\documentclass[twocolumn,preprintnumbers,floatfix,aps,prd,amssymb,floatfix,nofootinbib,balancelastpage,superscriptaddress,amsmath]{revtex4-1}
\pdfoutput=1
\usepackage[utf8]{inputenc}
\usepackage{amssymb}
\usepackage{amsmath}
\usepackage{amsfonts}
\usepackage{graphicx}
\usepackage{color}
\usepackage{xspace}
\usepackage{comment}
\usepackage{hyperref}
\usepackage[normalem]{ulem}
\usepackage[section]{placeins}
\usepackage{afterpage}
\usepackage{slashed}
\usepackage{enumerate}
\usepackage{enumitem}
\usepackage{caption}
\usepackage{subfigure}
\usepackage{float}
\usepackage{ulem}
\usepackage{verbatim}
\usepackage{multirow,rotating}
\usepackage[dvipsnames]{xcolor}
\usepackage{braket}
\usepackage{silence}
\definecolor{dcolour}{rgb}{.5, .5, .5}
\def\gsim{\raise0.3ex\hbox{$\;>$\kern-0.75em\raise-1.1ex\hbox{$\sim\;$}}}
\def\lsim{\raise0.3ex\hbox{$\;<$\kern-0.75em\raise-1.1ex\hbox{$\sim\;$}}}
\def\gsim{\raise0.3ex\hbox{$\;>$\kern-0.75em\raise-1.1ex\hbox{$\sim\;$}}}
\def\lsim{\raise0.3ex\hbox{$\;<$\kern-0.75em\raise-1.1ex\hbox{$\sim\;$}}}

\newcommand{\ba}[1]{\begin{eqnarray} \label{(#1)}}
\newcommand{\ea}{\end{eqnarray}}

\begin{document}

\title{ The strong coupling $g_{X J/\psi\phi}$ of $X(4700) \to J/\psi \phi$ in the light-cone sum rules }
\author{\textsc{Yiling Xie}}
\author{\textsc{Dazhuang He}}
\affiliation{Institute of Theoretical Physics, School of Physics, Dalian University of Technology, \\ 
No.2 Linggong Road, Dalian, Liaoning, 116024, People’s Republic of China}
\author{\textsc{Xuan Luo}}
\affiliation{School of Physics and Optoelectronics Engineering, Anhui University, \\
Hefei, Anhui 230601, People’s Republic of China}
\author{\textsc{Hao Sun}}
\email{haosun@dlut.edu.cn}
\affiliation{Institute of Theoretical Physics, School of Physics, Dalian University of Technology, \\ 
No.2 Linggong Road, Dalian, Liaoning, 116024, People’s Republic of China}

\begin{abstract}  

We assign the scalar tetraquark and the D-wave tetraquark state for $X(4700)$ 
and calculate the width of the decay $X(4700)$ $\to J/\psi \phi$ within the framework of light-cone sum rules. 
The strong coupling $g_{X J/\psi \phi}$ is obtained by considering the technique of soft-meson approximation.
We also investigate the mass and the decay constant of $X(4700)$ in the framework of SVZ sum rules. 
Our prediction for the mass is in agreement with the experimental measurement.
For the decay width of $X(4700)$ $\to J/\psi \phi$,
if we assigns $X$(4700) as a scalar $cs\bar{c}\bar{s}$ tetraquark state,
we obtain $\Gamma=(109^{+35}_{-24})$ MeV, which indicates that $X(4700)$ $\to J/\psi \phi$ is a predominant decay channel.
On the contrary, if $X$(4700) is assigned as a D-wave tetraquark state,
we obtain $\Gamma=(17.1^{+6.2}_{-4.0})$ MeV, so that $X$(4700) $\to$ $J/\psi \phi$ becomes a much smaller decay channel.
%And for the decay width of $X(4700)$ $\to J/\psi \phi$,
%if we assigns $X$(4700) as a scalar $cs\bar{c}\bar{s}$ tetraquark state,
%\added{the result will turn into $(109^{+35}_{-24})\ \text{MeV}$,
%and assigns $X$(4700) as a D-wave tetraquark state,
%the result will become  $(17.1^{+6.2}_{-4.0})\ \text{MeV}$ which is much smaller.}
%\deleted{$X$(4700) $\to$ $J/\psi \phi$ will the predominant process, 
%and that if one assign $X$(4700) as a D-wave tetraquark state,
%$X$(4700) $\to$ $J/\psi \phi$ will be a small decay channel. 
%Therefore the result imply that there exist multiple decays that not yet found in experiments.}
%\deleted{the possibility that $X(4700)$ could be a scalar tetraquark state if $X(4700)$ $\to J/\psi \phi$ is the predominant decay channel, or a D-wave tetraquark state
%if $X(4700)$ $\to J/\psi \phi$ is not the predominant one and there exist other decays.}

%\vspace{0.5cm}
\end{abstract}
%%%%%%%%%%%%%%%%%%%%%%%%%%%%%%%%%%%%%%%%%%%%%%%%%%%%%%%%%%%%%%%%%%%%%%
%\arxivnumber{}
%\pacs{}
\vskip10mm
\maketitle
\flushbottom
%%%%%%%%%%%%%%%%%%%%%%%%%%%%%%%%%%%%%%%%%%%%%%%%%%%%%%%%%%%%%%%%%%%%%%
\tableofcontents
\setcounter{footnote}{0}

\section{Introduction}
\label{I}

In 2016, the LHCb Collaboration analyzed the $B^{+}$ $\to J/\psi\phi K^+$ decay with $3\ \text{fb}^{-1}$ data of $pp$ collision at $\sqrt{s}=7$ and $8$ TeV \cite{LHCb:2016axx,LHCb:2016nsl}, confirmed there are four resonances in the $J/\psi \phi$ mass spectrum, i.e., $X(4140)$, $X(4274)$, $X(4500)$ and $X(4700)$. The spin parity number of $X(4140)$ and $X(4274)$ states are determined to be $J^{PC}=1^{++}$ with $5.7\sigma$ and $5.8\sigma$ significance,
respectively. The $J^{PC}$ of $X(4500)$ and $X(4700)$ states are $0^{++}$ with $5.2\sigma$ and $4.9\sigma$ significance, respectively. Their masses and widths are \cite{LHCb:2016axx}
\begin{equation}
   \begin{aligned}
   X(4140):&\text{M}=4146.5\pm4.5^{+4.5}_{-2.8}&\text{MeV},\\
           &\Gamma=83\pm21^{+21}_{-14}&\text{MeV},\\
   X(4274):&\text{M}=4273.3\pm8.3^{+17.2}_{-3.6}&\text{MeV},\\
           &\Gamma=56\pm11^{+8}_{-11}&\text{MeV},\\
   X(4500):&\text{M}=4506\pm11^{+12}_{-15}&\text{MeV},\\
           &\Gamma=92\pm21^{+21}_{-20}&\text{MeV},\\
   X(4700):&\text{M}=4704\pm10^{+14}_{-24}&\text{MeV},\\
           &\Gamma=120\pm31^{+42}_{-33}&\text{MeV}.\\
   \end{aligned}
\end{equation}

Since these resonance-like peaks appear in the $J/\psi\phi$ invariant mass spectrum, and $J/\psi\phi$ contains a $c\bar{c}$ pair and a $s\bar{s}$ pair, which implies that these states may be charmonium. The predicted mass of $\chi_{c0}(6P)$ is about $4669$ MeV in the screened potential (SP) model \cite{Gui:2018rvv}, which is very close to that of $X(4700)$. But its predicted total width is only about 16 MeV, too narrow to be comparable with $120\pm31^{+42}_{-33}$ MeV, the observed width  of $X(4700)$. Such discrepancy makes it difficult to understand the structure of $X(4700)$ in this way. 
In the constituent quark model, the authors in Refs.\cite{Fernandez:2016bqr,Ortega:2016hde} 
calculated the quark-antiquark spectrum for $J^{PC}=0^{++}$ channels, and showed that $X(4700)$ could appear as conventional charmonium states with quantum numbers $5^3P_0$ 
since it lies in the predicted mass and width ranges for $5^3P_0$. According to Non-Relativistic QCD (NRQCD) results, $X(4700)$ is also compatible with a charmonium $\chi_{c0}(4P)$\cite{Oncala:2017hop}. However, the higher charmonium states like $\chi_{cJ}(nP)$ would have numerous decay modes to open charm mesons, unfortunately there is no relevant mass spectrum appears in the $B\to$ $D^{(*)}_{(s)}\bar{D}^*_{(s)}K$ decays. In addition, the coupling of the higher charmonium states to $J/\psi\phi$ and to $J/\psi\omega$ should be very similar, but the BaBar collaboration measured the $J/\psi\omega$ mass spectrum in $B^{+}$ $\to J/\psi\omega K^+$ decay and did not find any structures resembling the $J/\psi\phi$ mass peaks \cite{LHCb:2016axx,LHCb:2016nsl}, which contradicts a charmonium interpretation for the $X(4700)$.
 
Though $X(4700)$ can't be a pure charmonium, it could be explained as a hybrid state, a charmonium state contains excited gluon fields. According to NRQCD, the mass of $0^{++}$ hybrid $1p_0(H_3)$ state is $4566$ MeV which is close to the spectrum of $X(4700)$. However, the charmonium fraction of $0^{++}$ hybrid is very small. So it is difficult to understand $X(4700)$ observed in the $J/\psi\phi$ channel since $ J/\psi $ mainly contains $c\bar{c}$ pair.

Besides hybrid, charmonium state can also be mixed with a light hadron to form a bound states, called hadrocharmonium. Considering that $X(4700)$ is observed in the $J/\psi\phi$ spectrum, it could be a bound state of $J/\psi-\phi$ or $\Psi(2S)-\phi$. However, the $J/\psi-\phi$ potential was found too weak, in lattice QCD, to form a bound states \cite{Ozaki:2012ce}. Furthermore, the $\Psi(2S)-\phi$ bound state has already been assigned to $X(4274)$ \cite{Panteleeva:2018ijz} so that rules out the possibility that $X(4700)$ be a s-wave $\Psi(2S)-\phi$ bound state. Anyway, $X(4700)$ could be other hadrocharmonium states in specific regions of the parameter space. 
Such explanation may require different binding mechanisms \cite{Panteleeva:2018ijz}.

To extrapolate the intrinsic structure of $X(4700)$, a myriad of theoretical studies has turned to the $cs\bar{c}\bar{s}$ tetraquark state. 

%\deleted{$X(4700)$ has been assigned as the 2S excited $cs\bar{c}\bar{s}$ tetraquark state that suggested by Maiani et al.\cite{Maiani:2016wlq},
%and within the relativized quark model in Ref.\cite{Lu:2016cwr}.}
It was suggested in Ref.\cite{Maiani:2016wlq,Lu:2016cwr} that $X(4700)$ can be assigned as the 2S excited $cs\bar{c}\bar{s}$ tetraquark state.
%To extrapolate the intrinsic structure of $X(4700)$, a myriad of theoretical studies has turned to the $cs\bar{c}\bar{s}$ tetraquark state. $X(4700)$ has been assigned as the 2S excited $cs\bar{c}\bar{s}$ tetraquark state that suggested by Maiani et al. in Ref.\cite{Maiani:2016wlq}, and that suggested in Ref.\cite{Lu:2016cwr} within the relativized quark model. 
$X$(4700) could also be explained as the radial excitation tetraquark state, for instance, of the hidden charm tetraquarks with quark content $\frac{1}{\sqrt{6}}(u\bar{u}+d\bar{d}-2s\bar{s})c\bar{c}$ in a diquark-antidiquark model \cite{Zhu:2016arf},
or the S-wave radial excited compact tetraquark states within the framework of the color flux-tube model with a multibody confinement potential \cite{Deng:2017xlb}.
In addition, $X$(4700) could be assigned as an first radial excitation of $X$(4350) tetraquark state through the color-magnetic interaction model \cite{Wu:2016gas}, or a 2S radial excited compact tetraquark state with $J^{PC}=0^{++}$ in the chiral quark model \cite{Yang:2019dxd}.
%$X$(4700) could also be explained as the radial excitation tetraquark state. For instance, $X$(4700) explained as the radial excitation of the hidden charm tetraquarks with quark content $\frac{1}{\sqrt{6}}(u\bar{u}+d\bar{d}-2s\bar{s})c\bar{c}$ based on the investigation in a diquark-antidiquark model \cite{Zhu:2016arf}, or the S-wave radial excited compact tetraquark states $[cs][\bar{c}\bar{s}]$ within the framework of the color flux-tube model with a multibody confinement potential \cite{Deng:2017xlb}. In addition, $X$(4700) can also be assigned as an orbital or radial excitation of tetraquark state $X$(4350) through the color-magnetic interaction model \cite{Wu:2016gas}, and a 2S radial excited compact tetraquark state with $J^{PC}=0^{++}$ in the chiral quark model \cite{Yang:2019dxd}.

Quite apart from the above excited tetraquark states explanation for $X(4700)$, there are several others. For example, it has been considered as a $0^{++}$ axial-vector diquark-antidiquark bound states in the diquark model \cite{Anwar:2018sol}, or a compact tetraquark state with $IJ^P = 00^+$ in the framework of the quark delocalization color screening model \cite{Liu:2021xje}. In particular, it can also be treated as a D-wave $cs\bar{c}\bar{s}$ tetraquark states \cite{Chen:2016oma} or a ground tetraquark state \cite{Wang:2016gxp} by using the sum rule approach develop by Shifman, Vainshtein and Zakharov (SVZ sum rules) \cite{Reinders:1984sr}.

Although the above theories have claimed the resonance peak in the $J/\psi \phi$ mass spectrum corresponds to the genuine resonance, there are other opinions\cite{Nakamura:2021bvs}. In Ref.\cite{Liu:2016onn,Ge:2021sdq}, the authors have investigated the open-charmed mesons, $J/\psi K^{*+}$, $\psi^\prime K^+$, and $\psi^\prime\phi$ re-scattering effects or threshold cusps in the process $B^+\to J/\psi\phi K^+$, in which $X$(4700) can be simulated by the $\psi^\prime\phi$ re-scattering via the $\psi^\prime K_1$ loops. Anyway, within the available experimental data, none of these theoretical interpretations can be completely accepted or excluded to the nature of $X$(4700). At this point, the structure of $X$(4700) is not yet fully settled.

As we mentioned, with the method of SVZ sum rules, $X$(4700) has been investigated in Ref.\cite{Chen:2016oma} and Ref.\cite{Wang:2016gxp}
in which its mass is predicted. 
%As we mention before, in ref.\cite{Chen:2016oma} and ref.\cite{Wang:2016gxp} investigated $X$(4700) by using the QCD sum rule. They calculate not the decay width but the mass of $X$(4700).
In this paper, we follow the same assumption that $X$(4700) is a D-wave tetraquark state and a groud tetraquark state simultaneously. We evaluate the mass of $X$(4700) in SVZ sum rules.
The results are compared with the prediction in Ref.\cite{Chen:2016oma, Wang:2016gxp}, and with the values in PDG \cite{ParticleDataGroup:2020ssz} to ensure the credibility of our calculation.
We then extend to evaluate the decay constant of $X$(4700), which is needed in the numerical calculation of the strong coupling $g_{X J/\psi\phi }$. While calculating $g_{X J/\psi\phi}$, the interpolating currents are still taken from Ref.\cite{Chen:2016oma, Wang:2016gxp}, and the method of light-cone sum rules (LCSR) is used and the full technical details are presented, in particular, the technique of soft-meson approximation is considered.
Based on these calculations, we finally obtain the decay width of $X\text{(4700)}\to$ $J/\psi\phi$. The results are compared with the the total decay widths of $X$(4700) in experimental measurements. Our present study can be regarded as a supplement to other previous works.

Our paper is organized as follows: In Section.\ref{II} the strong coupling $g_{XJ/\psi \phi}$ will be derived with the approach of light-cone sum rules. And we also calculate the mass and decay constant of the $X$(4700) state within two-point SVZ sum rule approach. The numerical results and discussions are shown in Section.\ref{III}. We reach our summary in Section.\ref{sec:summary}.

\section{Calculation Framework}
\label{II}

\subsection{ The strong coupling $g_{X J/\psi \phi}$ in the LCSR}

Before starting to predict the width of $X(\,4700\,)$ $\to J/\psi\phi$, we need to calculate the strong coupling $g_{X J/\psi\phi}$, in the framework of QCD LCSR. 
Let's strat by defining the two-point correlation function:
\begin{equation}\label{3}
	\begin{aligned}
	&\Pi_{\text{I},\mu}^{\text{LC}}(\,p+q,q\,)\\
        &=i\int \mbox{d}^4xe^{ipx}\braket{\phi(\,q\,)|\,T\{\,J_{\mu}^{J/\psi}(\,x\,)J_{\text{I}}^{X\dagger}(\,0\,)\,\}\,|0},
	\end{aligned}
\end{equation}
where $p$, $q$ are the four-momentum for $J/\psi$ and $\phi$ respectively and $\text{I}=(1),(2)$. Therefore $X(4700)$ has four-momentum $p^\prime=p+q$ according to the momentum conservation.
$J_{\mu}^{J/\psi}$ is the interpolating current of $J/\psi$,
and $J_{(1)}^{X}$, $J_{(2)}^{X}$ are that of $X(4700)$ in two different structures. They are given by \cite{Albuquerque:2008up, Becirevic:2013bsa}
\begin{equation}
	\begin{aligned}
		J_{\mu}^{J/\psi}(x)=&\bar{c}_i(x)\gamma_\mu c_i(x),\\
		J_{(1)}^{X}(x)=&\varepsilon_{ijk}\varepsilon_{imn}[s^T_j(x)C\gamma_\mu\gamma_5c_k(x)][\bar{s}_m(x)\gamma_5\gamma^\mu C\bar{c}_n^T(x)],\\
		J_{(2)}^{X}(x)=&c_a^T(x)C\gamma_{\mu 1}[D_{\mu 3}D_{\mu 4}s_b(x)]\\
		&(\bar{c}_a(x)\gamma_{\mu 2}Cs_b^T(x)+\bar{c}_b(x)\gamma_{\mu 2}C\bar{s}_a^T(x))\\
		\times&(g^{\mu 1\mu 3}g^{\mu 2\mu 4}+g^{\mu 1\mu 4}g^{\mu 2\mu 3}-g^{\mu 1\mu 2}g^{\mu 3\mu 4}/2)
	\end{aligned}
\end{equation}
where $a,b,i,j,k,m,n$ are the color indexes and $C$ is the charge conjugation matrix.

Next we need to establish a relation
between the correlation function $\Pi_{\text{I},\mu}^{\text{LC}}(p^\prime,q)$
and the strong coupling $g_{X J/\psi \phi}$. For a general dispersion relation, we have
\begin{equation}\label{dispersion}
	\begin{aligned}
		\Pi_{\text{I},\mu}^{\text{LC}}(p^\prime,q)=\frac{1}{\pi^2}\int\int \frac{ds_1ds_2\text{Im}\Pi_{\text{I},\mu}^{\text{LC}}(s_1,s_2)}{(s_1-p^2)(s_2-p^{\prime2})}+\cdots
	\end{aligned}
\end{equation}
where the subtraction terms and the single dispersion integrals are not shown, all of them will vanish after applying the double Borel transformation to Eq.\eqref{dispersion}.
Inserting in Eq.\eqref{3} with two complete sets of hadronic states and using Eq.\eqref{dispersion},
we obtain the phenomenological expression of the correlation function noted by $\Pi^{\text{LC,phen}}_{\text{I},\mu}(p^\prime,q)$
\begin{equation}\label{pheno}
	\begin{aligned}
		&\Pi^{\text{LC,phen}}_{\text{I},\mu}(p^\prime,q)=\\
		&\frac{\braket{0|J_{\mu}^{J/\psi}|J/\psi(p)}
		       \braket{\phi(q)J/\psi(p)|X(p^\prime)}
			 \braket{X(p^\prime)|J_{\text{I}}^{{X}\dagger}|0}}
			 {(p^{\prime2}-m_{\text{I},X}^2)(p^2-m_{J/\psi}^2)}\\
			&+\int_{s_1^{0\prime}}^\infty\int_{s_2^{0\prime}}^\infty
			   \frac{ds_1ds_2\ \rho_{\text{I},\mu}^{\text{LC,phen}}(s_1,s_2)}{(s_1-p^2)(s_2-p^{\prime2})}
			   +\int_{s_1^{0\prime}}^\infty\frac{ds_1\ \rho_{\text{I},1,\mu}^{\text{LC,phen}}(s_1)}{(s_1-p^2)}\\
			   &+\int_{s_2^{0\prime}}^\infty\frac{ds_2\ \rho_{\text{I},2,\mu}^{\text{LC,phen}}(s_2)}{(s_2-p^{\prime2})}.
	\end{aligned}
\end{equation}
Here contributions of the higher resonances and the continuum states are denoted by
$\rho_{\text{I},\mu}^{\text{LC,phen}}(s_1,s_2)$,
$s_1^{0\prime}$ and $s_2^{0\prime}$ denote the lowest thresholds of continuum states. $\rho_{\text{I},1,\mu}^{\text{phen}}(s_1)$
and $\rho_{\text{I},2,\mu}^{\text{phen}}(s_2)$ are the additional contributions to make the double dispersion integral finite\cite{Belyaev:1994zk}.
 The strong coupling $g_{XJ/\psi \phi}$ is defined
as an invariant constant parameterizing the hadronic matrix element
\begin{equation}\label{coupling}
	\begin{aligned}
		\braket{\phi(q)J/\psi(p)|X(p^\prime)}=&g_{XJ/\psi \phi}\\
		\times &[-(q\cdot p)(\varepsilon^*\cdot \varepsilon^\prime)+(q\cdot \varepsilon^*)(p\cdot\varepsilon^\prime)],
	\end{aligned}
\end{equation}
and the decay constants here are defined as:
\begin{equation}\label{decay constants}
	\begin{aligned}
		\braket{X(p^\prime)|J_{I}^{X\dagger}|0}&=m_{I,X}\, f_{I,X},\\
		\braket{0|J_{\mu}^{J/\psi}|J/\psi(p)}&=m_{J/\psi}f_{J/\psi}\, \varepsilon_\mu,
	\end{aligned}
\end{equation}
where $\varepsilon$, $\varepsilon^\prime$ are the polarization vectors of
the $J/\psi$ and $\phi$ respectively, $m_{J/\psi(X)}$ and $f_{J/\psi(X)}$ are the mass and decay constants of $J/\psi$($X(4700)$).

By inserting Eq.\eqref{coupling} and \eqref{decay constants} back to Eq.\eqref{pheno}, and performing the polarization sum, we can derive
\begin{equation}
	\begin{aligned}\label{1}
		\Pi^{\text{LC,phen}}_{\text{I},\mu}(p^\prime,q)&=\frac{m_{J/\psi}f_{J/\psi}\ m_{\text{I,X}} f_{\text{I,X}}\ g_{\text{XJ}/\psi \phi}}{(p^{\prime2}-m_{\text{I,X}}^2)(p^2-m_{J/\psi}^2)}\\
		&\times [ (p\cdot q)\ \epsilon_{\mu}^\prime-(p\cdot\epsilon^\prime)\ q_\mu ] +\cdots\\
		&=\Pi^{\text{LC,phen}}_{\text{I}}(p^\prime,q) [(p\cdot q)\epsilon_{\mu}^\prime-p\cdot\epsilon^\prime q_\mu ],
	\end{aligned}
\end{equation}
where we choose the structure proportional to $ \epsilon_{\mu}^\prime $ to work with.
The relevant form can be written as
\begin{equation}
	\begin{aligned}
		\Pi^{\text{LC,phen}}_{\text{I}}(p^\prime,q)&=\frac{m_{J/\psi}m_{\text{I},X}f_{J/\psi}f_{\text{I},X}g_{XJ/\psi \phi}}{(p^{\prime2}-m_{\text{I},X}^2)(p^2-m_{J/\psi}^2)}\\
		&+\int_{s_1^{0\prime}}^\infty\int_{s_2^{0\prime}}^\infty\frac{\mbox{d}s_1\mbox{d}s_2\rho^{\text{LC,phen}}_\text{I}(s_1,s_2)}{(s_1-p^2)(s_2-p^{\prime2})}+\cdots.
	\end{aligned}
\end{equation}
Applying the Borel transformations on variables $p^2$ and $p^{\prime2}=(p+q)^2$
to the correlation function yields
\begin{equation}\label{ha}
	\begin{aligned}
		&\mathcal{B}_{p^2}(M_1^2)\mathcal{B}_{p^{\prime 2}}(M_2^2)\Pi^{\text{LC,phen}}_{\text{I}}(p^\prime,q)=\\
		&m_{J/\psi}m_{\text{I},X}f_{J/\psi}f_{\text{I},X}g_{XJ/\psi \phi}\exp[-\frac{m_{J/\psi}^2}{M_1^2}-\frac{m_{\text{I},X}^2}{M_2^2}]\\
		&+\int_{s_1^{0\prime}}^\infty\int_{s_2^{0\prime}}^\infty\mbox{d}s_1\mbox{d}s_2 \exp[-\frac{s_1}{M_1^2}-\frac{s_2}{M_2^2}] \rho^{\text{LC,phen}}_{\text{I}}(s_1,s_2).
	\end{aligned}
\end{equation}

To proceed, we represent the OPE result for the correlation function
in the form of the double dispersion integral
\begin{equation}
	\begin{aligned}
	\Pi_{\text{I}}^{\text{LC,OPE}}(p^\prime,q)
	=\int_{s_1^{\prime}}^\infty\int_{s_2^{\prime}}^\infty \frac{ds_1ds_2\ \rho_{\text{I}}^{\text{LC,OPE}}(s_1,s_2)}{(s_1-p^2)(s_2-p^{\prime2})}
	+\cdots
	\end{aligned}
\end{equation}
with the double spectral density
\begin{equation}
	\begin{aligned}
	\rho_{\text{I}}^{\text{LC,OPE}}(s_1,s_2)=\frac{\text{Im}\Pi_{\text{I}}^{\text{LC,OPE}}(s_1,s_2)}{\pi^2}.
	\end{aligned}
\end{equation}
By choosing the structure proportional to $ \epsilon_{\mu}^\prime $
and performing the Borel transformations on variables $p^2$ and $p^{\prime2}=(p+q)^2$,
we find out
\begin{equation}\label{ope}
	\begin{aligned}
	\Pi^{\text{LC,OPE}}_\text{I}&(M_1^2,M_2^2)=\\
	&\int_{s_1^{\prime}}^\infty\int_{s_2^{\prime}}^\infty\mbox{d}s_1\mbox{d}s_2 \exp[-\frac{s_1}{M_1^2}-\frac{s_2}{M_2^2}]
	\rho^{\text{LC,OPE}}_\text{I}(s_1,s_2).
	\end{aligned}
\end{equation}

Then we employ the quark-hadron duality: assume the integral of the hadronic spectral density
$\rho^{\text{h}}_\text{I}(s_1,s_2)$ in the region $\{s_1\geq s_1^{0\prime},s_2\geq s_2^{0\prime}\}$
is equal to the integral of the OPE spectral density $\rho^{\text{LC,OPE}}_\text{I}(s_1,s_2)$
in a certain region of $\{s_1\geq s_1^{0},s_2\geq s_2^{0}\}$
\begin{equation}\label{hope}
	\begin{aligned}
		&\int_{s_1^{0\prime}}^\infty\int_{s_2^{0\prime}}^\infty\mbox{d}s_1\mbox{d}s_2 \exp[-\frac{s_1}{M_1^2}-\frac{s_2}{M_2^2}] \rho^{\text{LC,phen}}_\text{I}(s_1,s_2)=\\
		&\int_{s_1^{0}}^\infty\int_{s_2^{0}}^\infty\mbox{d}s_1\mbox{d}s_2 \exp[-\frac{s_1}{M_1^2}-\frac{s_2}{M_2^2}] \rho^{\text{LC,OPE}}_\text{I}(s_1,s_2).
	\end{aligned}
\end{equation}
Equating the double dispersion \eqref{ha} and \eqref{ope},
and substituting \eqref{hope} to \eqref{ha},
the result of LCSR for the strong coupling reads:
\begin{equation}\label{h}
	\begin{aligned}
	&g_{XJ/\psi \phi}=\frac{1}{m_{J/\psi}m_{\text{I},X}f_{J/\psi}f_{\text{I},X}}\exp[\,\frac{m_{J/\psi}^2}{M_1^2}
	+\frac{m_{\text{I},X}^2}{M_2^2}\,]\\
	&\times\int^{s_1^0}_{s_1^\prime}\int^{s_2^0}_{s_2^\prime}\mbox{d}s_1\mbox{d}s_2
	\exp[\,-\frac{s_1}{M_1^2}-\frac{s_2}{M_2^2}\,] \rho^{\text{LC,OPE}}_\text{I}(\,s_1,s_2\,).
	\end{aligned}
\end{equation}

Nevertheless, our situation differs from the standard one. 
From Eq.\eqref{3}, we see that the interpolating currents of $X(\,4700\,)$ is located at the space-time point $x=0$,
and the interpolating current of $J/\psi$ is located at the point $x$.
Therefore by contracting the $\bar{c}$ and $c$ quark fields, there remain two light quarks $s$ and $\bar{s}$
sandwiched between the $\phi$ state and the vacuum states, i.e., $\braket{\phi(\,q\,)|\,[\,\bar{s}(\,0\,)s(\,0\,)]\,|0}$.
We encounter the situation that the correlation function depends not on $\braket{\phi(\,q\,)|\,[\,\bar{s}(\,x\,)s(\,0\,)\,]\,|0}$ 
but $\braket{\phi(\,q\,)|\,[\,\bar{s}(\,0\,)s(\,0\,)\,]\,|0}$, 
so that the $\phi$ distribution amplitude disappears and reduces to normalization factor.
Such situation is possible to appear in the kinematical limit $q\rightarrow 0$ which means $(\,p+q\,)=p$, 
and the correlation function depends only on one variable $p^2$.
%\red{Here we adopt this approach, and following Ref.\cite{Belyaev:1994zk} refer to
Here, following Ref.\cite{Belyaev:1994zk}, 
we adopt the approach of soft-meson approximation, 
taking the limit $q\rightarrow 0$ and dealing with the double pole terms.

The approximation of $q\rightarrow 0$ simplifies the hadronic side of the sum rules,
but leads to a more complicated expression on its hadronic representation.
The ground state depends only on the variable $p^2$:
\begin{equation}
	\begin{aligned}
		&\Pi^{\text{LC,phen}}_\text{I}(\,p\,)=\frac{m_{J/\psi}m_{\text{I},X}f_{J/\psi}f_{\text{I},X}}{(\,p^2-m^2_\text{I,X}\,)^2} g_{XJ/\psi \phi}
			   +\cdots,
	\end{aligned}
\end{equation}
where $m^2_\text{I}=\frac{m_{J/\psi}^2+m_{\text{I},X}^2}{2}$ and the Borel transformation on the variable $p^2$ applied to this correlation function yields\cite{Belyaev:1994zk}
\begin{equation}
	\begin{aligned}
		 &\Pi^{\text{LC,phen}}_\text{I}(\,p^\prime,q\,)\\
                 &=\frac{1}{M^2}(\,m_{J/\psi}m_{\text{I},X}f_{J/\psi}f_{\text{I},X}g_{XJ/\psi \phi}+AM^2\,)e^{\frac{-m^2_\text{I}}{M^2}}+C,
	\end{aligned}
\end{equation}
where the constant A incorporates all unsuppressed contributions, and C contains all exponentially suppressed contributions.
To remove unsuppressed contributions, we perform the operator that the script in\cite{Ioffe:1983ju}
\begin{equation}
	\begin{aligned}
		(\,1-M^2\frac{d}{dM^2}\,)M^2e^{m_\text{I}^2/M^2}
	\end{aligned}
\end{equation}
on both sides of the sum rules expression: the phenomenological side and the OPE side.
So Eq.\eqref{h} becomes
\begin{equation}\label{couplingn}
	\begin{aligned}
		&g_{XJ/\psi \phi}=\frac{1}{m_{J/\psi}m_{\text{I},X}f_{J/\psi}f_{\text{I},X}}(\,1-M^2\frac{d}{dM^2}\,)M^2\\
		&\times\int^{\hat{s}^\prime}_{\hat{s}_0}\mbox{d}\hat{s}\exp[\,\frac{m_{J/\psi}^2}{2M^2}+\frac{m_{\text{I},X}^2}{2M^2}-\frac{\hat{s}}{M^2}\,] \rho^{\text{OPE}}_\text{I}(\,\hat{s}\,).
	\end{aligned}
\end{equation}
Because of soft-meson approximation, the continuum state depends not on two variables $s_1$ and $s_2$ but one that relabel as $\hat{s}$.

\subsection{OPE side calculation}

We already know that $g_{XJ//psi /phi}$ is related to the OPE part of the correlation function, so we are going to calculate it.
%Firstly, one need to determine the correlation function by using quark propagators and light-cone distribution amplitudes (LCDAs) of $\phi$.
According to the Wick Theorem, we can obtain
\begin{equation}
	\begin{aligned}
		&\braket{\phi(\,q\,)|\, \text{T}\, \{\, J_{\mu}^{J/\psi}(\,x\,)J_{(1)}^{\text{X}\dagger}(\,0\,)\,\}\,|0}\\
		&=\braket{\phi(q)|\, [\, \bar{s}^a_\alpha(\,0\,) s^d_\beta(\,0\,)\, ]\, |0} \\
		&\times[\, \gamma_5\tilde{S}^{ib}_c(\,x\,)\gamma_\mu\tilde{S}^{ie}_c(\,-x\,)\gamma_\nu\gamma_5 \\
		&-\gamma_\nu\gamma_5\tilde{S}^{ib}_c(\,x\,)\gamma_\mu \tilde{S}^{ie}_c(\,-x\,)\gamma_5\, ]_{\alpha\beta}.\\
	\end{aligned}
\end{equation}
and
\begin{equation}
	\begin{aligned}
	&\braket{\phi(\,q\,)|\, \text{T}\{\, J_{\mu}^{J/\psi}(\,x\,)J_{(2)}^{\text{X}\dagger}(\,0\,)\, \}\, |0}\\
		&=\braket{\phi(\,q\,)|\, [\, (\,\bar{s}^b\overleftarrow{D}_{\mu 4}\overleftarrow{D}_{\mu 3}\,)_\alpha(\,0\,) s^b_\beta(\,0\,)\, ]\, |0}\\
	&\times[\, \gamma_{\mu 1}\tilde{S}^{ac}_c(\,x\,)\gamma_\nu\tilde{S}^{ac}_c(\,-x\,)\gamma_{\mu 2}\, ]_{\alpha\beta}\\
	&+\braket{\phi(\,q\,)|\ [\, (\,\bar{s}^b\overleftarrow{D}_{\mu 4}\overleftarrow{D}_{\mu 3}\,)_\alpha(\,0\,) s^a_\beta(\,0\,)\, ]\, |0}\\
	&\times[\, \gamma_{\mu 1}\tilde{S}^{ac}_c(\,x\,)\gamma_\nu\tilde{S}^{bc}_c(\,-x\,)\gamma_{\mu 2}\, ]_{\alpha\beta}\ .
	\end{aligned}
\end{equation}
Therefore, the correlation functions become
\begin{equation}\label{2}
	\begin{aligned}
	\Pi_{(1),\mu}^{\text{OPE}}(\,p^\prime,q\,)
	&=i\int d^4xe^{ipx}\braket{\phi(q)|\, \text{T}\{\, J_{\mu}^{J/\psi}(\,x\,)J^{\text{X}\dagger}_{(1)}(\,0\,)\, \}\, |0}\\
	&=i\int d^4xe^{ipx}\varepsilon_{ijk}\varepsilon_{imn}\braket{\phi(\,q\,)|\, [\ \bar{s}^j_\alpha(\,0\,) s^m_\beta(\,0\,)\, ]\, |0}\\
	&\quad\times[\, \gamma_\mu\gamma_5\tilde{S}^{kl}_c(\,x\,)\gamma_\nu\tilde{S}^{nl}_c(\,-x\,)\gamma_\nu\gamma_5\, ]_{\alpha\beta},
	\end{aligned}
\end{equation}
and
\begin{equation}\label{31}
	\begin{aligned}
	&\Pi_{(2),\mu}^{\text{OPE}}(\,p^\prime,q\,)
	=i\int d^4xe^{ipx}\braket{\phi(\,q\,)|\, \text{T}\{\,J_{\mu}^{J/\psi}(\,x\,)J^{\text{X}\dagger}_{(2)}(\,0\,)\,\}\, |0}\\
	&=i\int d^4xe^{ipx}\{\braket{\phi(\,q\,)|\ [\,(\bar{s}^b(\,0\,)\overleftarrow{D}_{\mu 4}\overleftarrow{D}_{\mu 3})_\alpha s^b_\beta(\,0\,)\,]\, |0}\\
	&\times[\, \gamma_{\mu 1}\tilde{S}^{ac}_c(x)\gamma_\nu\tilde{S}^{ac}_c(-x)\gamma_{\mu 2}\ ]_{\alpha\beta}\\
	&+\braket{\phi(\,q\,)|\, [\, (\,\bar{s}^b(\,0\,)\overleftarrow{D}_{\mu 4}\overleftarrow{D}_{\mu 3}\,)_\alpha s^a_\beta(\,0\,)\, ]\, |0}\\
	&\times[\, \gamma_{\mu 1}\tilde{S}^{ac}_c(\,x\,)\gamma_\nu\tilde{S}^{bc}_c(\,-x\,)\gamma_{\mu 2}\, ]_{\alpha\beta}\},
	\end{aligned}
\end{equation}
where we introduce the notation
\begin{equation}
	\begin{aligned}
	\tilde{S}_q(\,x\,)=CS^T_{q}(\,x\,)C,	 		
	\end{aligned}
\end{equation}
and $S_q(x)$ is the propagator of quark $q$.
%, $C$ is the charge conjugation operator.
For the heavy quark propagator on the light-cone we employ
its expression in terms of \cite{Reinders:1984sr}
\begin{equation}\label{Hpropagator}
\begin{aligned}
S^{ab}_{c}(x)&=\\
&i\int\frac{d^4k}{(2\pi)^4}e^{-ikx}
[\frac{\delta_{ab}(\slashed{k}+m_c)}{k^2-m^2_c} \\
&-\frac{g_sG_{ab}^{\alpha\beta}}{4}\frac{\sigma_{\alpha\beta}(\slashed{k}+m_c)+(\slashed{k}+m_c)\sigma_{\alpha\beta}}{(k^2-m_c^2)^2}]\\
&+\frac{g_sD_\alpha G^n_{\beta\lambda}t^n_{ij}(f^{\lambda\beta\alpha}+f^{\lambda\alpha\beta})}{3(k^2-m_c^2)^4}\\
&-\frac{g_s^2(t^at^b)_{ij}G^a_{\alpha\beta}G^b_{\mu\nu}(f^{\alpha\beta\mu\nu}+f^{\alpha\mu\beta\nu}+f^{\alpha\mu\nu\beta})}{4(k^2-m_c^2)^5},
\end{aligned}
\end{equation}
with
\begin{equation}
\begin{aligned}
&f^{\lambda\alpha\beta}=(\slashed{k}+m_c)\gamma_\lambda(\slashed{k}+m_c)\gamma_\alpha(\slashed{k}+m_c)\gamma_\beta(\slashed{k}+m_c),\\
&f^{\alpha\beta\mu\nu}=(\slashed{k}+m_c)\gamma_\alpha(\slashed{k}+m_c)\gamma_\beta(\slashed{k}+m_c)\\
&~~~~~~~~~~~~~~~~\gamma_\mu(\slashed{k}+m_c)\gamma_\nu(\slashed{k}+m_c),
\end{aligned}
\end{equation}
here we use the shorthand notation
\begin{equation}
\begin{aligned}
G_{ab}^{\mu\nu}\equiv G_i^{\mu\nu}t^i_{ab},\ \ \ i=1,2,\cdots,8.
\end{aligned}
\end{equation}
It is convenient to perform the summation over the
color indices by performing the replacement
\begin{equation}\label{summation}
\begin{aligned}
\bar{s}^d_\alpha(0) s^{d^\prime}_\beta(0)=\frac{1}{3}\delta_{dd^\prime}\bar{s}_\alpha(0) s_\beta(0),
\end{aligned}
\end{equation}
and using the expansion
\begin{equation}\label{expansion}
\begin{aligned}
\bar{s}_\alpha(0) s_\beta(0)\equiv\frac{1}{4}\Gamma^a_{\alpha\beta}\bar{s}(0)\Gamma^a s(0),
\end{aligned}
\end{equation}
where the sum runs over the Dirac structures $a$	
\begin{equation}
\begin{aligned}
\Gamma^a=1,\gamma_5,\gamma_\mu,i\gamma_5\gamma_\mu,\frac{\sigma_{\mu\nu}}{\sqrt{2}}.
\end{aligned}
\end{equation}
Substituting the summation Eq.\eqref{summation} and the expansion Eq.\eqref{expansion} 
into Eq.\eqref{2} and Eq.\eqref{31}, we obtain
\begin{equation}\label{opecorrelation}
\begin{aligned}
\Pi_{(1),\mu}(p^\prime,q)
	&=i\int d^4xe^{ipx}\braket{\phi(q)|T\{J_{\mu}^{J/\psi}(x)J^{X\dagger}_{(1)}(0)\}|0}\\
	&=i\int d^4xe^{ipx}\varepsilon_{ijk}\varepsilon_{imn}\braket{\phi(q)|[\bar{s}^j(0)\Gamma^a s^m(0)]|0}\\
	&\times\text{Tr}[\gamma_\mu\gamma_5\tilde{S}^{kl}_c(x)\gamma_\nu\tilde{S}^{nl}_c(-x)\gamma_\nu\gamma_5\Gamma^a],
\end{aligned}
\end{equation}
and
\begin{equation}\label{opecorrelation1}
	\begin{aligned}
	\Pi_{(2),\mu}(p^\prime,q)
	&=i\int d^4xe^{ipx}\braket{\phi(q)|T\{J_{\mu}^{J/\psi}(x)J^{X\dagger}_{(2)}(0)\}|0}\\
	&=i\int d^4xe^{ipx}\{\braket{\phi(q)|[(\bar{s}^b(0)\overleftarrow{D}_{\mu 4}\overleftarrow{D}_{\mu 3})\Gamma^e s^b(0)]|0}\\
	&\times\text{Tr}[\gamma_{\mu 1}\tilde{S}^{ac}_c(x)\gamma_\nu\tilde{S}^{ac}_c(-x)\gamma_{\mu 2}]\\
	&+\braket{\phi(q)|[(\bar{s}^b(0)\overleftarrow{D}_{\mu 4}\overleftarrow{D}_{\mu 3})\Gamma^e s^a(0)]|0}\\
	&\times\text{Tr}[\gamma_{\mu 1}\tilde{S}^{ac}_c(x)\gamma_\nu\tilde{S}^{bc}_c(-x)\gamma_{\mu 2}]\},
	\end{aligned}
\end{equation}

\begin{figure}[h!]
\centering
  \includegraphics[width=5cm]{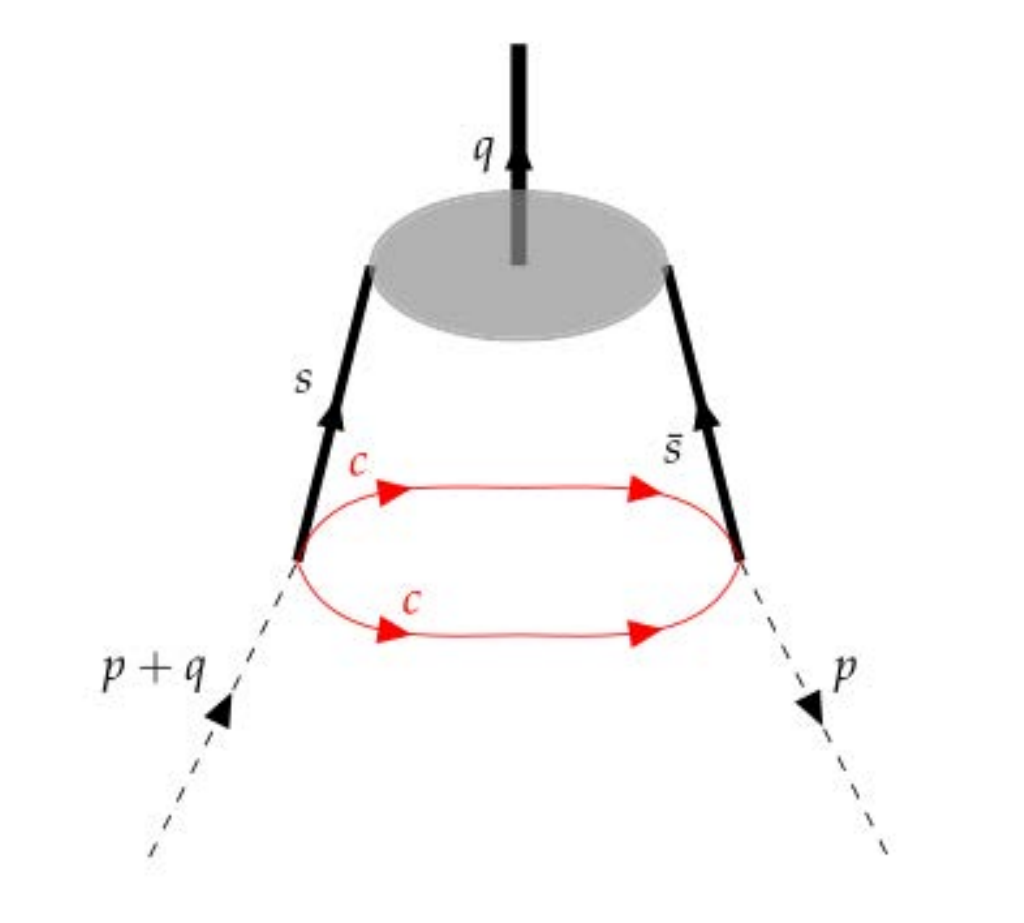}
  \includegraphics[width=5cm]{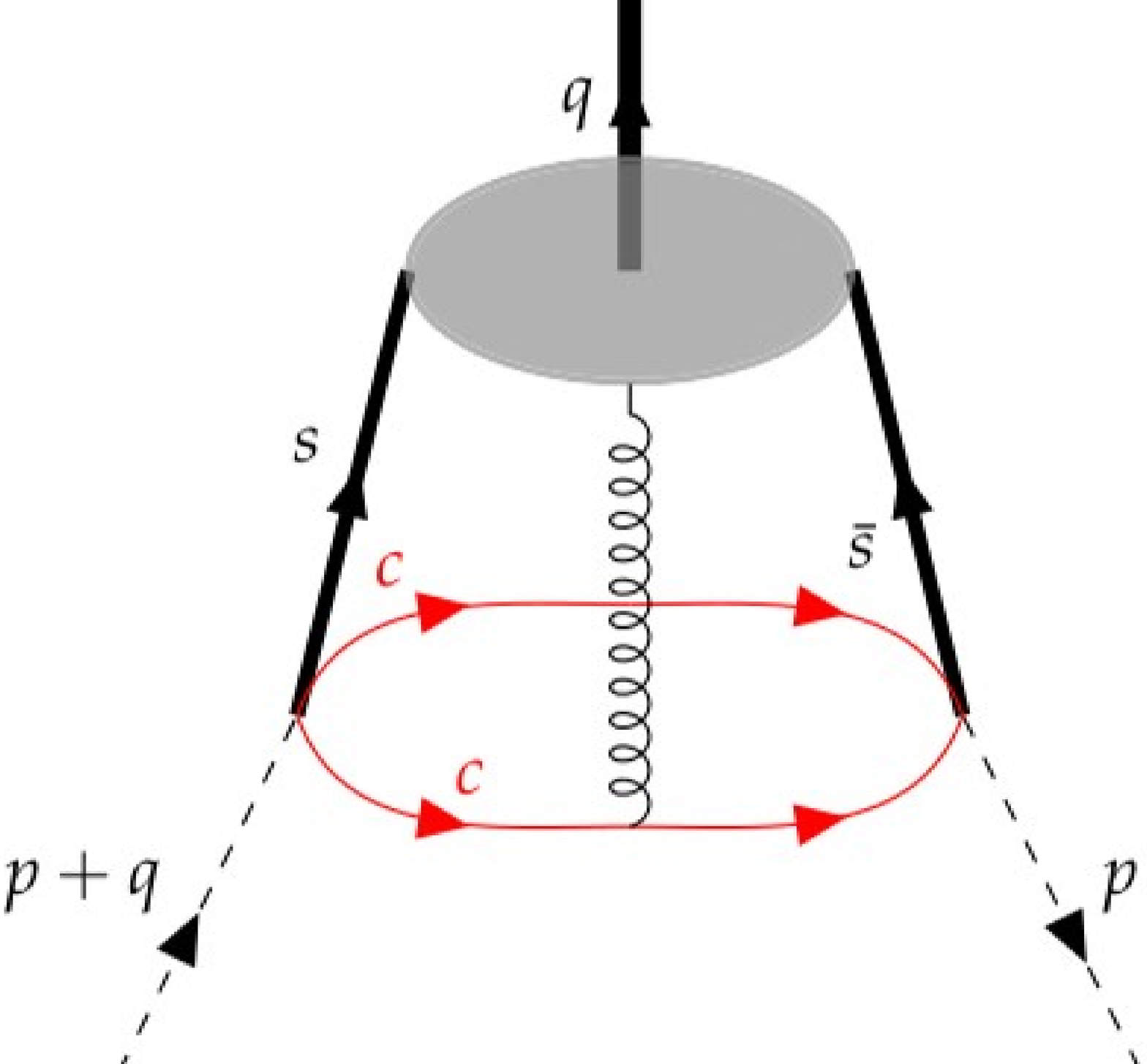}
  \includegraphics[width=5cm]{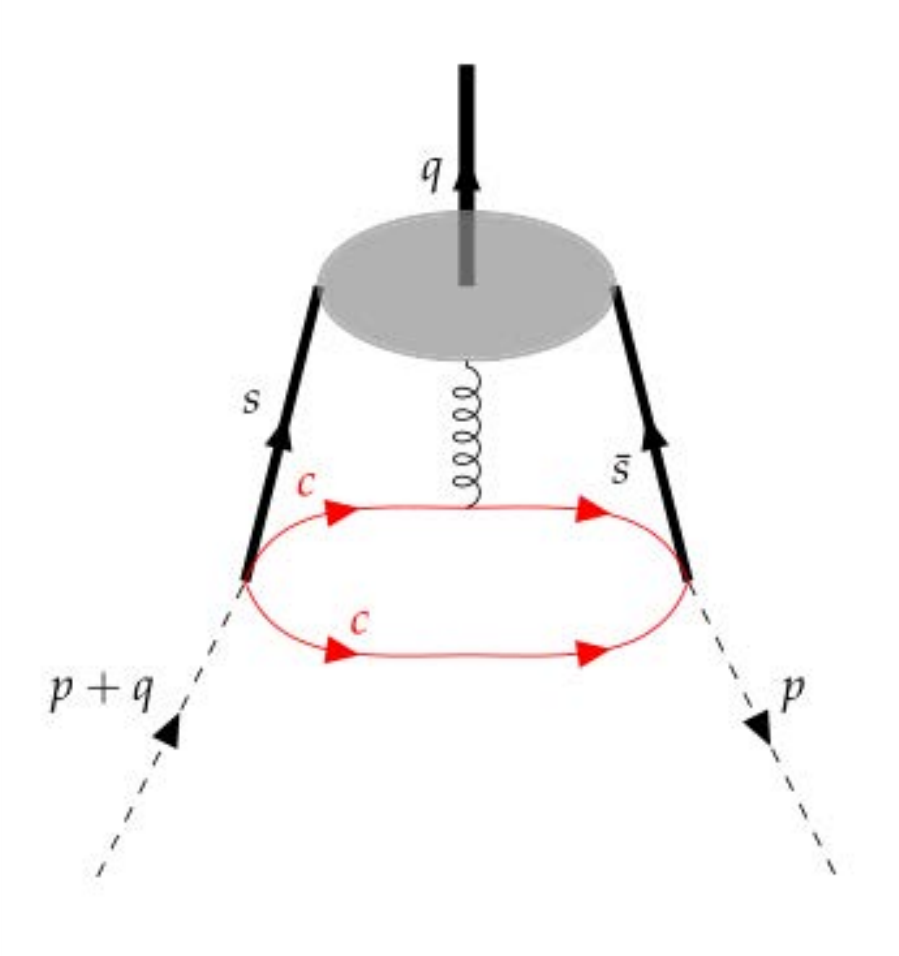}
  \caption{
  The leading order and one-gluon exchange diagrams contribute to $\Pi_{(1),\mu}(p^\prime,q)$, which are the main contribution.}
 \label{Fig:fig1}
\end{figure}
\begin{figure}[h!]
\centering
  \includegraphics[width=5cm]{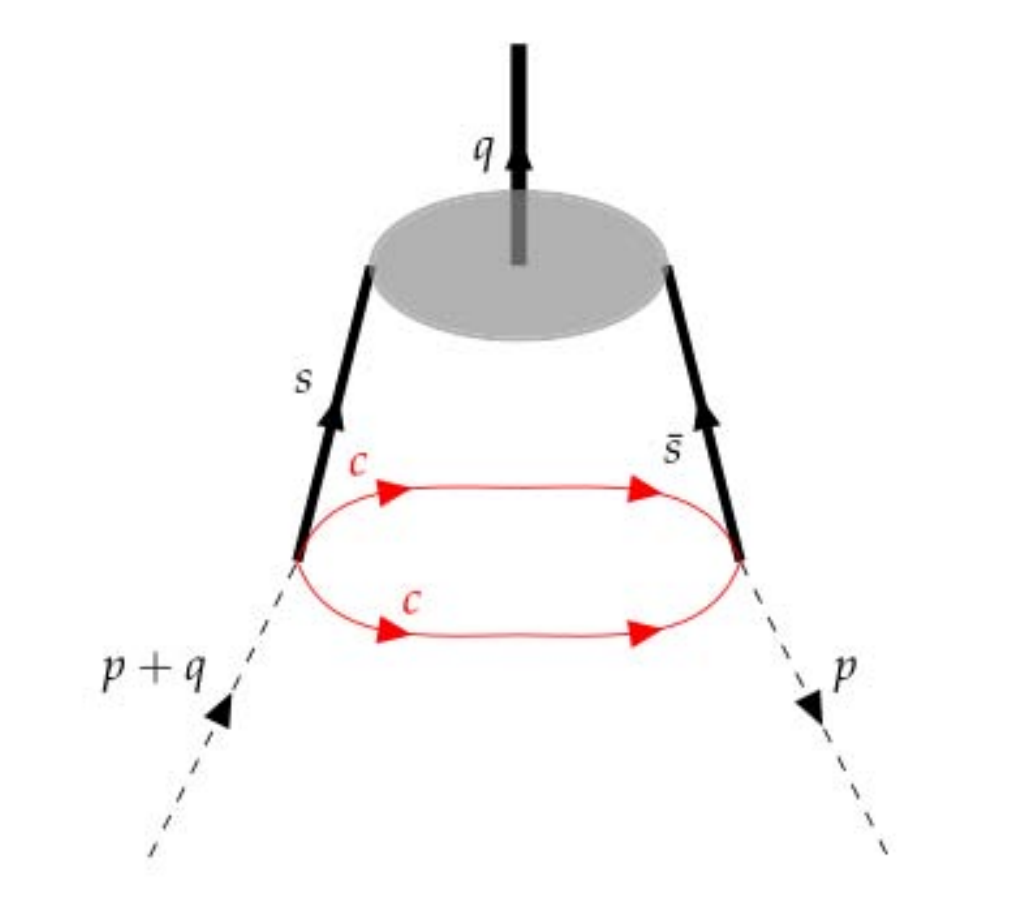}
  \caption{The leading order diagram contribute to $\Pi_{\mu (2)}(p^\prime,q)$. The one-gluon exchange vanish in $\Pi_{(2),\mu}(p^\prime,q)$.}
\label{Fig:fig2}
\end{figure}
The Feynman diagram of $\Pi_{(1),\mu}(p^\prime,q)$ is shown in FIG.\ref{Fig:fig1} and the Feynman diagrams of $\Pi_{(2),\mu}(p^\prime,q)$ is shown in FIG.\ref{Fig:fig2}. FIG.\ref{Fig:fig1} corresponds to the leading order and next-leading order contribution of $\Pi_{(1),\mu}(p^\prime,q)$. FIG.\ref{Fig:fig2} corresponds to the leading order contribution of $\Pi_{(2),\mu}(p^\prime,q)$, 
which is the dominant one compare to, for example, the one-gluon exchange contribution.

Now we can substitute the propagator in Eq.\eqref{opecorrelation}
by the perturbative term of Eq.\eqref{Hpropagator}. Using the Particle Distribution Amplitudes (DAs) of $\phi$ in Appendix \ref{appendix:A} 
and contract the color index by the SU(N) algebra
\begin{eqnarray}
%	\begin{aligned}
	\varepsilon_{abc}\varepsilon_{dec}\delta_{ad}\delta_{bi}\delta_{ei}=-C_A(1-C_A)=6,
%	\end{aligned}
\end{eqnarray}
we will encounter four-dimensional integrals in the momentum spaces. After above operation, some terms in Eq.\eqref{opecorrelation}
will proportional to, for example
\begin{equation}\label{four integral}
	\begin{aligned}
		\int\frac{d^4k_1}{(2\pi)^4}\int\frac{d^4k_2}{(2\pi)^4}\frac{e^{-i(k_1-k_2)x} k_1\cdot k_2}{(k_1^2-m^2_c)(k_2^2-m^2_c)}[(p\cdot q)\epsilon_{\mu}^\prime-p\cdot\epsilon^\prime q_\mu].
	\end{aligned}
\end{equation}
The main steps to calculate some four-dimensional integrals like \eqref{four integral} can be done by dimension regulation.
Choose the structure proportional to $ \epsilon_{\mu}^\prime$, we can derive the corresponding spectral density
\begin{equation}
	\begin{aligned}
		\rho^{\text{OPE}}_{(1)}(\hat{s})=&-\frac{f_\phi^{\|} m_{\phi} \sqrt{\hat{s} \left(\hat{s}-4 m^2\right)} \left(2 m^2+\hat{s}\right)}{12\pi ^2 \hat{s}}\\
		&-\frac{f_\phi^{\|} m_c^2 m_{\phi} \braket{\frac{\alpha_s}{\pi}GG} \left(6 m^2-\hat{s}\right) \sqrt{\hat{s} \left(\hat{s}-4 m^2\right)}}{144 \pi ^2 \hat{s}^2 \left(\hat{s}-4 m^2\right)^2}\\
		&\frac{-\zeta_{4\phi}^{\bot}f_\phi^{\bot}m_c m_\phi^2(2m_c^2-\hat{s})\sqrt{\hat{s}(\hat{s}-4m_c^2)}}{2\pi^2\hat{s}^2(4m_c^2-\hat{s})}\\
		&+\frac{\tilde{\zeta}_{4\phi}^{\bot}f_\phi^{\bot}m_cm_\phi^2\sqrt{\hat{s}(\hat{s}-4m_c^2)}}{2\pi^2\hat{s}(4m_c^2-\hat{s})},
	\end{aligned}
\end{equation}
and
\begin{equation}
	\begin{aligned}
		\rho^{\text{OPE}}_{(2)}(\hat{s})=\int_0^1d uu^2\phi_2^{\bot}(u)\frac{\sqrt{2}m_c m_{\phi}^2f_{\phi}^{\bot}\sqrt{\hat{s}(\hat{s}-4m_c^2)}}{2\pi^2\hat{s}},
	\end{aligned}
\end{equation}
where $m_{\phi}$ and $f_{\phi}^{\bot}$ are the mass and decay constant of $\phi$ respectively. $\phi_2^{\bot}(u)$ is LCDA. $\zeta_4^{\bot}$ and $\tilde{\zeta}_4^{\bot}$ are three-particle distribution parameters defined in Appendix \ref{appendix:A}, $m_c$ is charm-quark mass. The strong coupling is then evaluated by Eq.\eqref{couplingn}. In addition, we can easily obtain the decay width of $X(4700)$ $\to J/\psi \phi$ by applying the usual feynman diagram method \cite{Dias:2013xfa}
 \begin{equation}
 	\begin{aligned}
 		&\Gamma_\text{I}(X(4700)\to J/\psi \phi)=\frac{(g_{XJ/\psi \phi})^2}{24\pi m_{\text{I},X}^2}\\
 		&\times\lambda(m_{\text{I},X},m_{J/\psi},m_{\phi})\left(3+\frac{\lambda(m_{\text{I},X},m_{J/\psi},m_{\phi})}{m^2_{J/\psi}}\right),
 	\end{aligned}
 \end{equation}
where 
 \begin{equation}
 	\begin{aligned}
 		\lambda(a,b,c)=\frac{\sqrt{a^4+b^4+c^4-2*(a^2b^2+b^2c^2+c^2a^2)}}{2a}.
 	\end{aligned}
 \end{equation}

\subsection{ The mass and the decay constant of $X(4700)$ }

To calculate the mass and the decay constant, we start from the two-point correlation function:
\begin{equation}\label{corm}
	\begin{aligned}
	\Pi_{\text{I}}^{\text{SVZ}}(p)=i\int \mbox{d}^4xe^{ipx}
	\braket{0|T\{J_{\text{I}}^{X}(x)J_{\text{I}}^{X\dagger}(0)\}|0}.
	\end{aligned}
\end{equation}
%\deleted{where the interpolating currents are given by the following expression:}
% \begin{equation}
% 	\begin{aligned}
% J_{(1)}^{X\dagger}(x)=&\varepsilon_{ijk}\varepsilon_{imn}[\bar{c}_k(x)\gamma_5\gamma_\mu C\bar{s}_j^T(x)][c_n^T(x)C\gamma^\mu\gamma_5s_m(x)],\\
% 		J_{(2)}^{X\dagger}(x)=&[(\bar{s}_b(x)\overleftarrow{D}_{\mu 3}\overleftarrow{D}_{\mu 4})\gamma_{\mu 1}C\bar{c}_a^T]\\
% 		&(s_b^T(x)C\gamma_{\mu 2}Cc_a(x)\pm s_a^T(x)C\gamma_{\mu 2}Cc_b(x))\\
% 		\times&(g^{\mu 1\mu 3}g^{\mu 2\mu 4}+g^{\mu 1\mu 4}g^{\mu 2\mu 3}-g^{\mu 1\mu 2}g^{\mu 3\mu 4}/2).
% 	\end{aligned}
% \end{equation}
We first calculate the correlation function in terms of the phenomenological expression 
by inserting in Eq.\eqref{corm} with a complete set of hadronic states:
\begin{equation}
	\begin{aligned}
		\Pi^{\text{SVZ,phen}}_{\text{I}}(p)&=
		\frac{\braket{0|J_\text{I}^{X}|X(p)}
		\braket{X(p)|J_\text{I}^{X\dagger}|0}}{m_{\text{I},X}^2-p^2}\\
		&+\int_{s^\prime}^\infty d \hat{s}\frac{\rho_{I}^{\text{SVZ,phen}}(\hat{s})}{\hat{s}-p^2},
	\end{aligned}
\end{equation}
where $\rho_{\text{I}}^{\text{SVZ,phen}}(\hat{s})$ stands for 
the contributions of the higher resonances and the continuum states. 
The subtraction terms are not shown because they will vanish after the Borel transformation.
%\deleted{We define the decay constant $f_{\text{I},X}$ according to}
% \begin{equation}
% 	\begin{aligned}
% 		\braket{0|J_I^{X}|X(p)}=m_Xf_X^I.
% 	\end{aligned}
% \end{equation}
After performing polarization sum we can derive
\begin{equation}\label{POLE-CONT}
	\begin{aligned}
	\Pi_{\text{I}}^{\text{SVZ,phen}}(p)=\frac{m_{\text{I},X}^2f_{\text{I},X}^2}{m_{\text{I},X}^2-p^2}
	+\int_{s^\prime}^\infty d\hat{s}\frac{\rho_{\text{I}}^{\text{SVZ,phen}}(\hat{s})}{\hat{s}-p^2}.
	\end{aligned}
\end{equation}
As we can see, there is a pole appearing on the right-hand side of Eq.\eqref{POLE-CONT}. 
%\deleted{the first term showing a pole that contributes to the mass and decay constant.}
The way of removing the pole is to perform the Borel transformation on Eq.\eqref{POLE-CONT}, which yields
\begin{equation}
	\begin{aligned}
		&\Pi_{\text{I}}^{\text{SVZ,phen}}(M^2)=\\
		&m_{\text{I},X}^2f_{\text{I},X}^2e^{-m_{\text{I},X}^2/M^2}+\int_{s^\prime}^\infty d\hat{s}\rho_{\text{I}}^{\text{SVZ,phen}}(\hat{s})e^{-\hat{s}/M^2}.
	\end{aligned}
\end{equation}

Now we turn to consider the correlation function in the OPE side, 
after contracting the heavy and light quark in term of Wick Theorm, we obtain
\begin{equation}\label{propagator1}
	\begin{aligned}
		\Pi_{(1)}^{\text{SVZ,OPE}}(p)=&i\int \mbox{d}^4xe^{ipx}\epsilon_{ijk}\epsilon_{imn}\epsilon_{i^\prime j^\prime k^\prime}\epsilon_{i^\prime m^\prime n^\prime}\\
		&\text{Tr}[\tilde{S}^{jj^\prime}_s(x)\gamma_\mu\gamma_5S_c^{kk^\prime}(x)\gamma_5\gamma_\alpha]\\
		&\text{Tr}[S^{mm^\prime}_s(-x)\gamma_5\gamma_\mu\tilde{S}_c^{nn^\prime}(-x)\gamma_\alpha\gamma_5],
	\end{aligned}
\end{equation}
and
\begin{equation}\label{propagator2}
	\begin{aligned}
		&\Pi_{(2)}^{\text{SVZ,OPE}}(p)=i\int \mbox{d}^4xe^{ipx}\\
		&\text{Tr}[\gamma_{\mu 1}(\overrightarrow{D}_{\mu 3}\overrightarrow{D}_{\mu 4}S^{bb^\prime}_s(x-y)\overleftarrow{D}_{\nu 3}\overleftarrow{D}_{\nu 4})\gamma_{\nu 1}\gamma_5\tilde{S}_c^{aa^\prime}(-x)]|_{y=0}\\
		&(\text{Tr}[\gamma_{\mu 2}\tilde{S}^{bb^\prime}_s(-x)\gamma_{\nu 2}S_c^{aa^\prime}(-x)]\\
		&+\text{Tr}[\gamma_{\mu 2}\tilde{S}^{ba^\prime}_s(-x)\gamma_{\nu 2}S_c^{ab^\prime}(-x)]\\
		&-\text{Tr}[\gamma_{\mu 2}\tilde{S}^{ab^\prime}_s(-x)\gamma_{\nu 2}S_c^{ba^\prime}(-x)]\\
		&+\text{Tr}[\gamma_{\mu 2}\tilde{S}^{aa^\prime}_s(-x)\gamma_{\nu 2}S_c^{bb^\prime}(-x)]\\
		\times&(g^{\mu 1\mu 3}g^{\mu 2\mu 4}+g^{\mu 1\mu 4}g^{\mu 2\mu 3}-g^{\mu 1\mu 2}g^{\mu 3\mu 4}/2)\\
		\times&(g^{\nu 1\nu 3}g^{\nu 2\nu 4}+g^{\nu 1\nu 4}g^{\nu 2\nu 3}-g^{\nu 1\nu 2}g^{\nu 3\nu 4}/2).
	\end{aligned}
\end{equation}
where
\begin{equation}
	\begin{aligned}
		\tilde{S}_q^{ab}(x)=CS_q^{ab}(x)C.
	\end{aligned}
\end{equation}
For propagators of the $u$, $d$ and $s$ quarks in coordinate-space, 
An expression for propagators is as follows\cite{Huang:2010dc,Agaev:2016mjb}:
\begin{equation}
	\begin{aligned}
		S_{q,ab}(x)&=\frac{i\delta_{ab}\slashed{x}}{2\pi^2x^4}-\frac{\delta_{ab}m_q}{4\pi^2x^2}-\frac{{\langle \bar{q}q\rangle }}{12}\\
		&-\frac{i}{32\pi^2}\frac{\lambda^n}{2}g_sG_{\mu\nu}^n\frac{1}{x^2}(\sigma^{\mu\nu}\slashed{x}+\slashed{x}\sigma^{\mu\nu})\\
		&+\frac{i\delta_{ab}\slashed{x}m_q{\langle \bar{q}q\rangle }}{48}
		-\frac{\delta_{ab}{\langle \bar{q}g_s\sigma Gq \rangle}x^2}{192}\\
		&+\frac{i\delta_{ab}x^2\slashed{x}m_q\langle\bar{q}g_s\sigma Gq\rangle}{1152}\\
		&-\frac{i\delta_{ab}x^2\slashed{x}g_s^2\langle\bar{q}q\rangle^2}{7776}-\frac{\delta_{ab}x^4\langle\bar{q}q\rangle\langle g_s^2GG\rangle}{27648}.
	\end{aligned}
\end{equation}
Notice the situation here is different from LCSR, where we do not encounter the light quark propagator expressed in the coordinate-space.
We have to face divergences in the double integrals like
\begin{equation}
	\begin{aligned}
		\int \frac{d^4x}{x^{2n}}\int\int\frac{d^4k_1d^4k_2e^{ipx-ik_1x-ik_2x}}{(k_1^2-m_c^2)(k_2^2-m_c^2)}.
	\end{aligned}
\end{equation}
As shown in Ref.\cite{Agaev:2016dev}, by using Fourier transformation
\begin{equation}
	\begin{aligned}
		\begin{aligned}
			&\frac{1}{\left(x^{2}\right)^{n}}=\int \frac{d^{D} p}{(2 \pi)^{D}} e^{-i p \cdot x} i(-1)^{n+1} 2^{D-2 n} \pi^{D / 2} \\
				&\times \frac{\Gamma(D / 2-n)}{\Gamma(n)}\left(-\frac{1}{p^{2}}\right)^{D / 2-n}
\end{aligned}
	\end{aligned}
\end{equation}
and perform dimension regulation\cite{Matheus:2006xi} and extract the imaginary part of results, we can obtain the results without any divergences.

The correlation function $\Pi^{\text{SVZ,OPE}}_{\text{I}}(p)$ can be represented as the dispersion integral
 \begin{equation}
 	\begin{aligned}
 		\Pi^{\text{SVZ,OPE}}_\text{I}(p)=\int_{4m_c^2}^\infty\mbox{d}\hat{s}\frac{\rho^{\text{SVZ,OPE}}_\text{I}(\hat{s})}{\hat{s}-p^2},
 	\end{aligned}
 \end{equation}
where $\rho^{\text{SVZ,OPE}}_\text{I}(\hat{s})$ is the corresponding spectral density.

By performing the Borel transformation on $\Pi^{\text{SVZ,OPE}}_\text{I}(p)$
and adopting the quark-hadron duality,
one can obtain
\begin{equation}
	\begin{aligned}
		&\Pi^{SVZ,OPE}_{\text{I}}(M^2,\infty)\equiv\int_{4m_c^2}^\infty\mbox{d}\hat{s}\Tilde{\rho}^{\text{SVZ,OPE}}_\text{I}(\hat{s})e^{-\hat{s}/M^2}\\
		&=m_{\text{I},X}^2f_{\text{I},X}^2e^{-m_{\text{I},X}^2/M^2}+\int_{s_0}^\infty\rho^{\text{SVZ,phen}}_\text{I}(\hat{s})e^{-\hat{s}/M^2}.
 	\end{aligned}
\end{equation}
Then subtract the continuum contribution yields:
\begin{equation}
	\begin{aligned}
		m_{\text{I},X}^2f_{\text{I},X}^2e^{-m_{\text{I},X}^2/M^2}&=\int_{4m_c^2}^{s_0}\mbox{d}\hat{s}\rho^{\text{SVZ,OPE}}_\text{I}(\hat{s})e^{-\hat{s}/M^2}\\
 	\end{aligned}
\end{equation}
The mass of the $X(4700)$ state can be evaluated from the sum rule
\begin{equation}
	\begin{aligned}
		m_{\text{I},X}^2=\frac{\int_{4m_c^2}^{s_0}\mbox{d}\hat{s}\hat{s}\rho^{\text{SVZ,OPE}}_\text{I}(\hat{s})e^{-\hat{s}/M^2}}{\int_{4m_c^2}^{s_0}\mbox{d}\hat{s}\rho^{\text{SVZ,OPE}}_\text{I}(\hat{s})e^{-\hat{s}/M^2}}.
	\end{aligned}
\end{equation}
where the spectral densities are  provided in APPENDIX \ref{appendix:B}.

\section{Numerical calculation}
\label{III}

\subsection{Input parameters}

In this section, we analyze the numerical results for the coupling constant and the decay width of $X(4700) \to J/\psi \phi$, and present the mass and decay constant of $X(4700)$ as well. We adopt the following parameters for the numerical calculation. The current charm-quark mass, $m_c=(1.275\pm 0.025)$ GeV, the $J/\psi$-meson mass $m_{J/\psi}=(3096.900\pm 0.006)$ MeV and $\phi$(980) mass $m_{\phi}=(990\pm 20)$ MeV 
from the Particla Data Group (PDG) \cite{ParticleDataGroup:2020ssz}. 
The $J/\psi$ and $\phi$(980) decay constants are taken as $f_{J/\psi}=0.405$ GeV \cite{Dias:2013xfa}, $f_{\phi}=0.18\pm 0.015$ GeV \cite{Colangelo:2010bg}. The current-quark-mass for the s-quark is $m_s=93^{+11}_{-5}$ MeV from PDG. The decay constants of $\phi$ is taken as $f_{\phi}^{\|}=0.215$ GeV. The Gegenbauer moments, $a_1^{\|}=a_1^{\bot}=0$ and $a_2^{\|}=0.18$, $a_2^{\bot}=0.14$ \cite{Ball:2007zt}. The parameters $\zeta_4^{\bot}$ and $\widetilde{\zeta}_4^{\bot}$ are taken as $\zeta_4^{\bot}=-0.01$ and $\widetilde{\zeta}_4^{\bot}=-0.03$ \cite{Ball:2007zt}.
In addition, we also need to know the values of the non-perturbative vacuum condensates. The related parameters are \cite{Hu:2021lkl}
\begin{equation}
	\begin{aligned}
		&\langle{\bar{q}q}\rangle=-(0.24\pm 0.01)^3\ \text{GeV}^3,\\
		&\braket{\bar{s}s}=(0.8\pm0.1)\times \braket{\bar{q}q},\\
		&\braket{\frac{\alpha_s}{\pi} GG}=(0.012)\ \text{GeV}^4,\\
		&\braket{g_s\bar{s}\sigma Gs}=m_0^2\times\braket{\bar{s}s},\\
		&m_0^2=0.8\ GeV^2,\\
		%&{\color{red}m_s(2GeV)}=95\pm5 \text{MeV},\\
		&m_c=1.275\pm0.025\ \text{MeV}.\\
	\end{aligned}
\end{equation}

The sum rule predictions for the mass, the decay constant and the coupling constant depend on two parameters: 
Borel mass $M^2$ and continuum threshold $s_0$. The value of $s_0$ is being correlated with the onset of excited states of $X(4700)$. But according to the experimental data, there is no resonance activity related to the first excited states of $X(4700)$. We should turn to annother way. From Table \ref{table:I} which entail the masses calculations of charmonia and bottomonia, we can find the mass discrepancy between the ground state and first excited state are all around $0.5$ GeV. Besides, the experimental data in Table \ref{table:II} which extract from PDG prove most of the calculations in Table \ref{table:I}.
Furthermore, one can refer to previous  QCD sum rules caculations that assign $X(4140)$ and $X(4685)$ as the 1S and 2S tetraquark states
respectively and that assign $Z_c(3900)$ and $Z_c(4430)$ as the 1S and 2S tetraquark states respectively and so on (see Table \ref{table:III}). The mass difference between the 1S and 2S tetraquark states are about $0.4\sim 0.6 $ GeV. So we accept
the mass discrepancy and employ
\begin{equation}
	\begin{aligned}
		(4.70+0.40)^2\ \text{GeV}^2\leq s_0\leq (4.70+0.60)^2\ \text{GeV}^2.
	\end{aligned}
\end{equation}
or rewrite it as
\begin{equation}
	\begin{aligned}
		(5.20-0.10)^2\ \text{GeV}^2\leq s_0\leq (5.20+0.10)^2\ \text{GeV}^2.
	\end{aligned}
\end{equation}

\begin{table}[H]
	\centering
	\caption{Quark model masses calculated for the first three levels of charmonia and bottomonia \cite{Olpak:2016wkf}.}
\begin{tabular}{|c|c|c|c|c|c|c|}
\hline 
Masses  &\multicolumn{3}{|c|}{$c\bar{c}$}  & \multicolumn{3}{|c|}{$b\bar{b}$}  \\
\hline \hline 
M(GeV) $\backslash$ n & n=1 & n=2 & n=3 & n=1 & n=2 & n=3 \\
\hline 
$M_{ ^{3}P_{0}}\left(\chi_{q 0}\right)$ & 3.37 & 3.88 & 4.30 & 9.81 & 10.2 & 10.7 \\
\hline 
$M_{ ^{3} P_{1}}\left(\chi_{q 1}\right)$ & 3.54 & 3.97 & 4.33 & 9.89 & 10.3 & 10.6 \\
\hline 
$M_{ ^{1} P_{1}}\left(h_{q}\right)$ & 3.53 & 3.96 & 4.37 & 9.88 & 10.3 & 10.6 \\
\hline 
$M_{ ^{3} P_{2}}\left(\chi_{q 2}\right)$ & 3.54 & 3.98 & 4.34 & 9.89 & 10.3 & 10.6\\
\hline
\end{tabular}
\label{table:I}
\end{table}

\begin{table}[H]
	\centering
	\caption{Masses of experimentally observed states in Particle Data Group listings\cite{ParticleDataGroup:2012pjm}.}
\begin{tabular}{|c|c|c|c|c|c|c|}
\hline Masses & \multicolumn{3}{|c|}{$c \bar{c}$} & \multicolumn{3}{|c|}{$b \bar{b}$} \\
\hline \hline$M(M e V) \backslash n$ & $n=1$ & $n=2$ & $n=3$ & $n=1$ & $n=2$ & $n=3$ \\
\hline$M_{ ^{3} P_{0}}\left(\chi_{q 0}\right)$ & $3414.75$ & $-$ & $-$ & $9859.44$ & $10232.5$ & $-$ \\
\hline$M_{ ^3 P_{1}}\left(\chi_{q 1}\right)$ & $3510.66$ & $-$ & $-$ & $9892.78$ & $10255.46$ & $10512.1$ \\
\hline$M_{ ^1P_{1}}\left(h_{q}\right)$ & $3525.38$ & $-$ & $-$ & $9899.3$ & $10259.8$ & $-$ \\
\hline$M_{ ^3 P_{2}}\left(\chi_{q 2}\right)$ & $3556.20$ & $3922.5$ & $-$ & $9912.21$ & $10268.65$ & $-$ \\
\hline
\end{tabular}
\label{table:II}
\end{table}

\begin{table}[H]
	\centering
	\caption{The mass difference between the 1S and 2S hidden-charm tetraquark states with the possible assignments \cite{Wang:2021lkg}.}
	\begin{tabular}{|c|c|c|c|c|}
	\hline \hline $J^{P C}$ & $1 \mathrm{~S}$ & 2 $\mathrm{~S}$ & $\text { Mass difference }$ & $\text { References }$ \\
	\hline $1^{++}$ & $X(4140)$ & $X(4685)$ & 566 $\mathrm{MeV}$ & {\cite{Wang:2021ghk,Wang:2018qpe}} \\
	\hline $1^{+-}$ & $Z_{c}(3900)$ & $Z_{c}(4430)$ & $591 \mathrm{MeV}$ & \cite{Maiani:2014aja,Nielsen:2014mva,Wang:2014vha}\\
	\hline $0^{++}$ & $X(3915)$ & $X(4500)$ & $588 \mathrm{MeV}$ & \cite{Lebed:2016yvr,Wang:2016gxp} \\
	\hline $1^{+-}$ & $Z_{c}(4020)$ & $Z_{c}(4600)$ & $576 \mathrm{MeV}$ & \cite{Chen:2019osl,Wang:2019hnw} \\
	\hline \hline
	\end{tabular}
\label{table:III}	
\end{table}

After fixing $s_0$, we use two extra criteria to constrain
the Borel mass $M^2$: 
\begin{enumerate}
\item
To obtain the minimal value of $M^2$, we require that the contribution of the
condensates like $\braket{\bar{q}g_sGq}$ and higher dimension condensates in the OPE is smaller than $5\%$ of the total contribution:
\begin{equation}
	\begin{aligned}
	\mathrm{CVG} \equiv\left|\frac{\Tilde{\Pi}_I^{\braket{\bar{q}g_sGq}+\cdots}\left( M^{2},\infty\right)}{\Tilde{\Pi}_I^{OPE}\left(M^{2},\infty \right)}\right| \leq 5\%,
	\end{aligned}
\end{equation}
where dots denote higher dimension contributions.
\item Ensure that the one-pole in Eq.\eqref{POLE-CONT} is valid, we require that the pole contribution (PC) should be larger than $30\%$ to determine the upper limit on $M^2$
\begin{equation}
	\begin{aligned}
		\text{PC}=\frac{\Tilde{\Pi}_I^{OPE}(M^2,s_0)}{\Tilde{\Pi}_I^{OPE}(M^2,\infty)}\ge  30\% .
	\end{aligned}
\end{equation}
\end{enumerate}

\begin{figure}[htbp]
\centering
  \includegraphics[width=7cm]{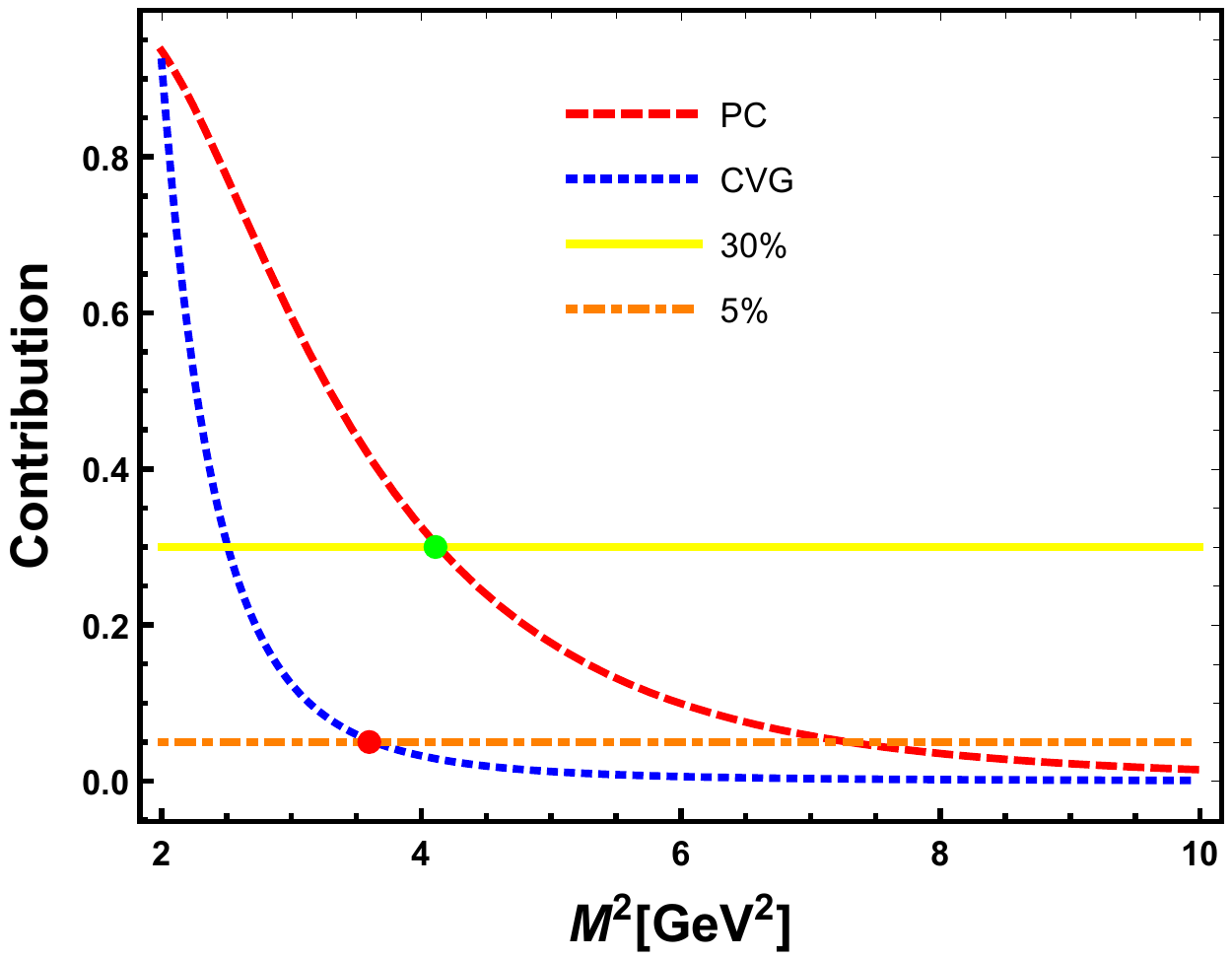}
  \caption{
  Convergence (CVG) and pole contribution (PC) for $J_{(1)}^{X}$. }
\label{Fig:fig3}
\end{figure}

\begin{figure}[htbp]
\centering
  \includegraphics[width=7cm]{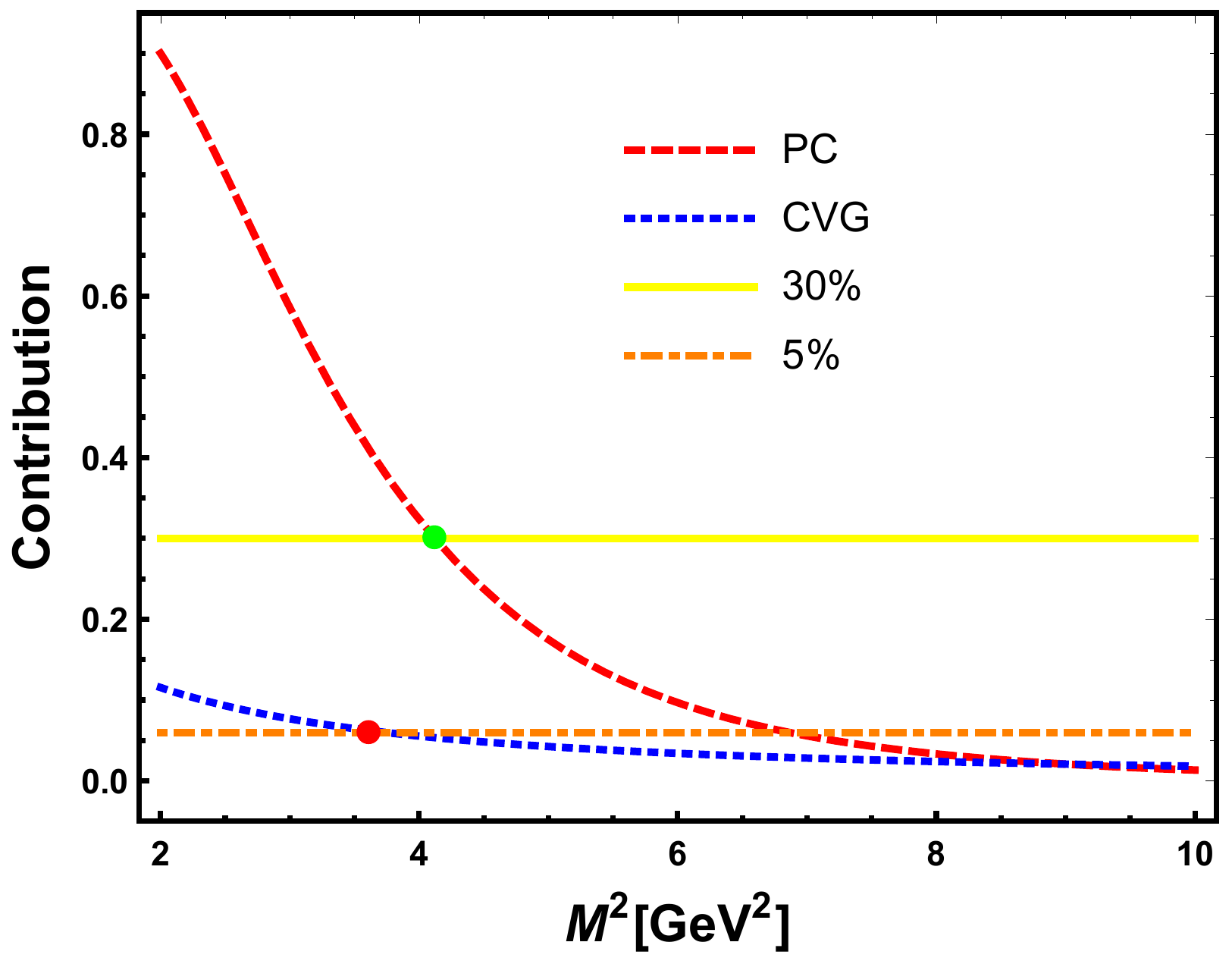}
  \caption{
  Convergence (CVG) and pole contribution (PC) for $J_{(2)}^{X}$. }
\label{Fig:fig4}
\end{figure}

As depicted in FIG.\ref{Fig:fig3} and FIG.\ref{Fig:fig4}, CVG and PC decrease as $M^2$ grows higher. The green dot betokens that CVG turns into $30\%$, where the maximal Borel mass $M^2$ can be obtained. And the red dot present PC converges with $5\%$ horizontal for $J_{(1)}^{X}$ and $J_{(2)}^{X}$, from which point, we can select the minimum Borel mass. Subsequently, we require the working region of the Borel parameter for mass and decay constant of $J_{(1)}^{X}$ to be in the region of
\begin{equation}
	\begin{aligned}
		3.60\ \text{GeV}^2\leq M^2\leq 4.13\ \text{GeV}^2,
	\end{aligned}
\end{equation}
and $J_{(2)}^{X}$ to be in that of
\begin{equation}
	\begin{aligned}
		3.63\ \text{GeV}^2\leq M^2\leq 4.17\ \text{GeV}^2.
	\end{aligned}
\end{equation}

\subsection{The mass, decay constant}

\begin{figure}[htbp]
\centering
  \includegraphics[width=7cm]{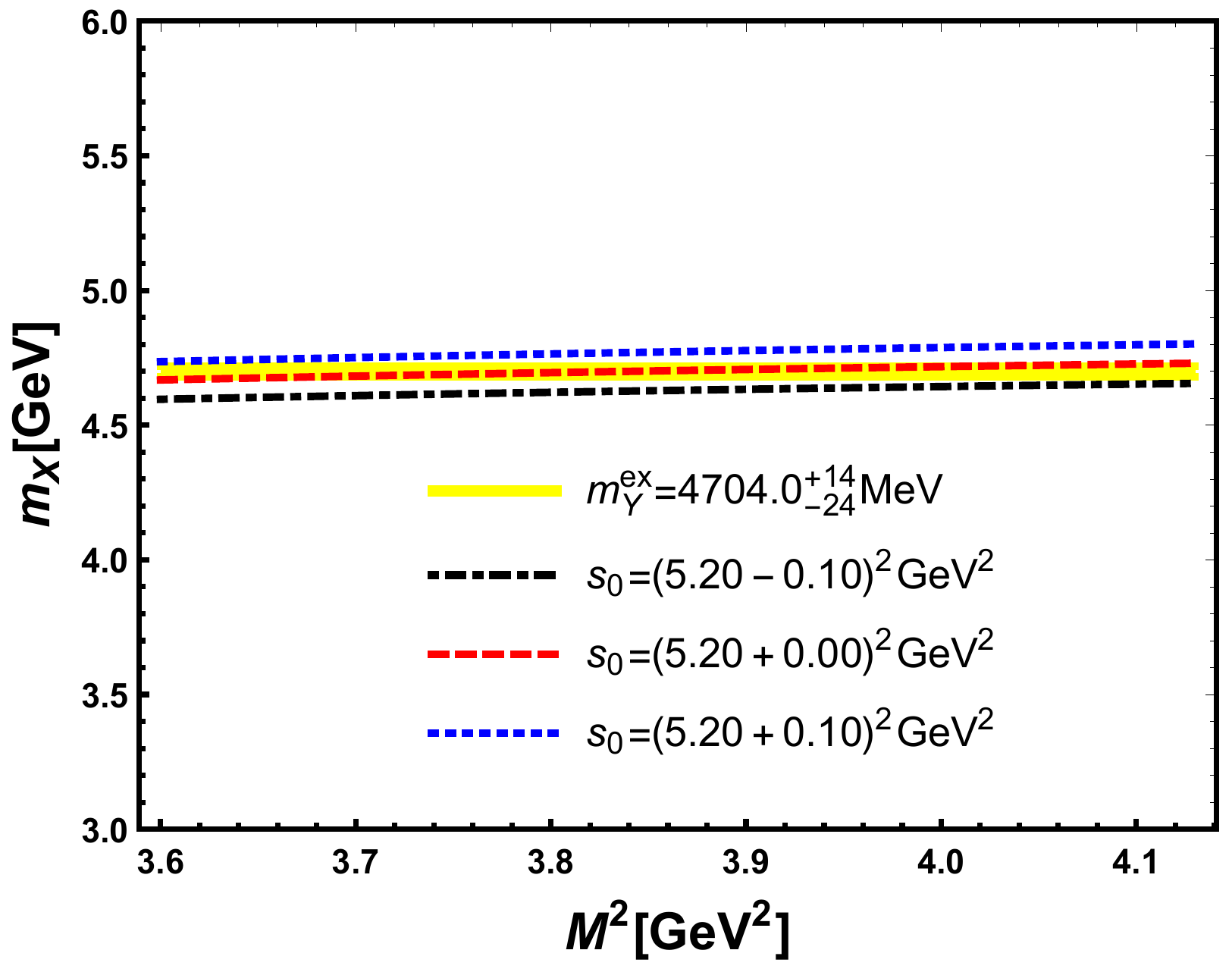}
  \includegraphics[width=7cm]{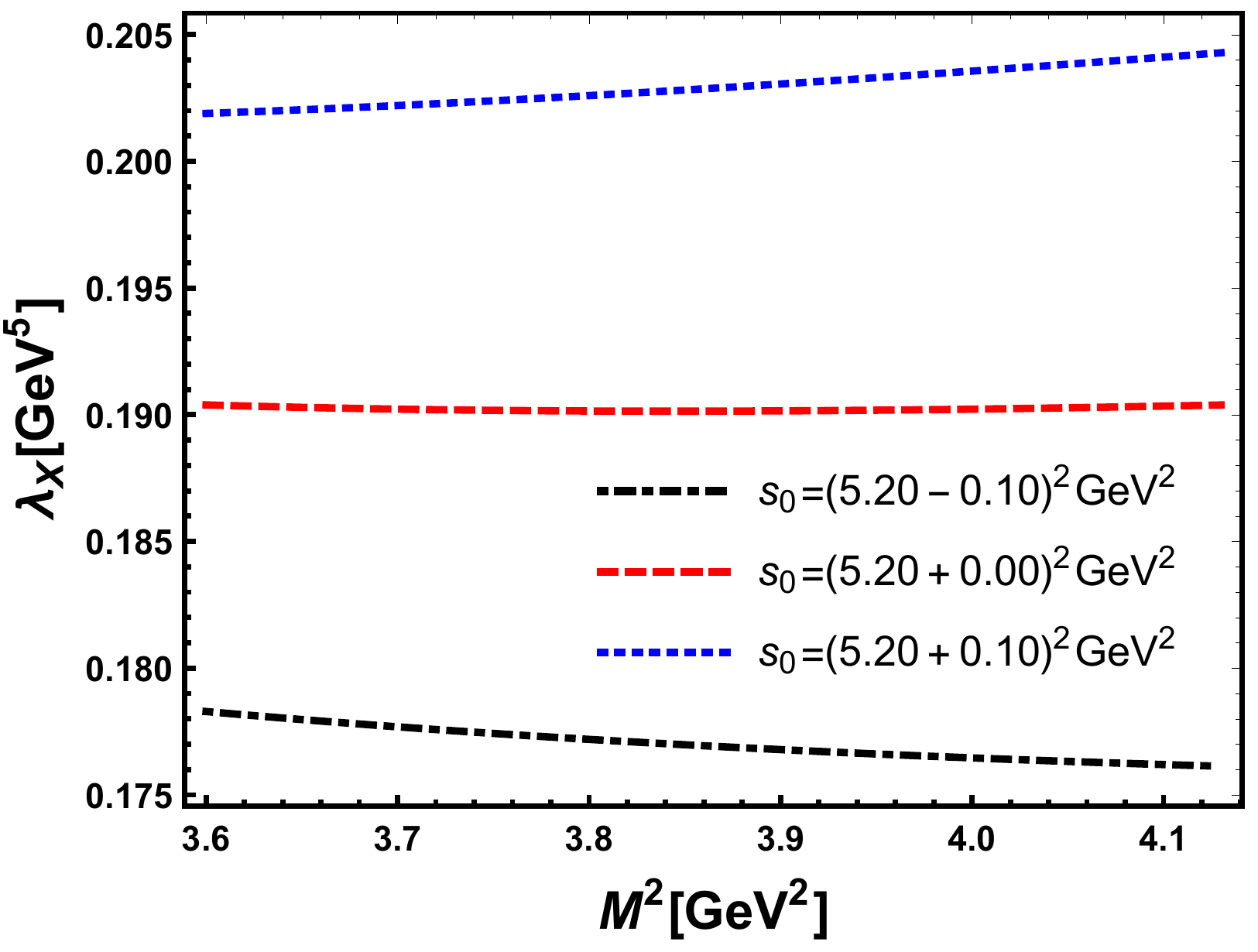}
  \caption{
 The mass [first] and the decay constant [second] of a scalar tetraquark state $X(4700)$ as a function of the Borel parameter $M^2$ at different fixed values of $s_0$.}
\label{Fig:fig5}
\end{figure}

\begin{figure}[htbp]
\centering
  \includegraphics[width=7cm]{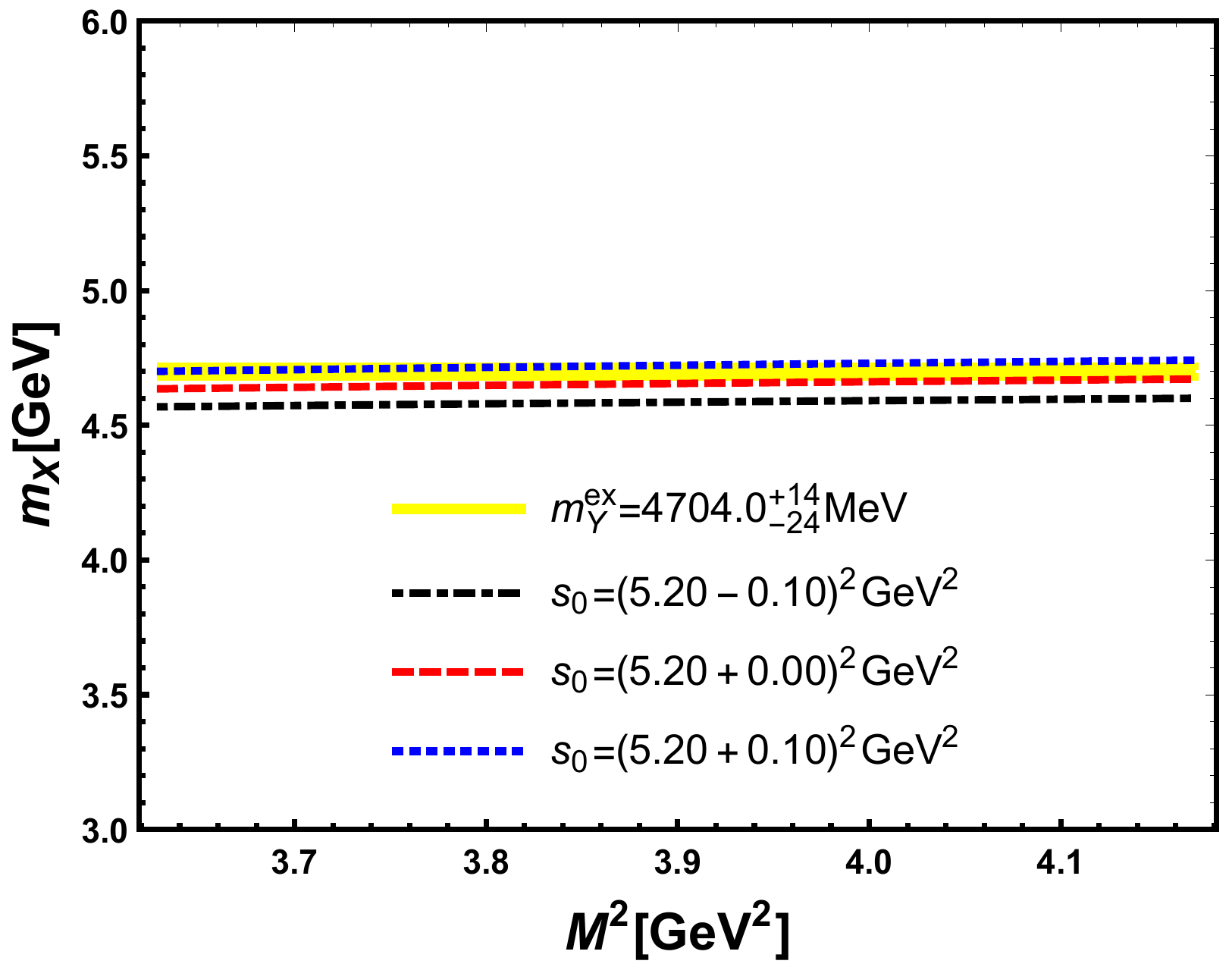}
  \includegraphics[width=7cm]{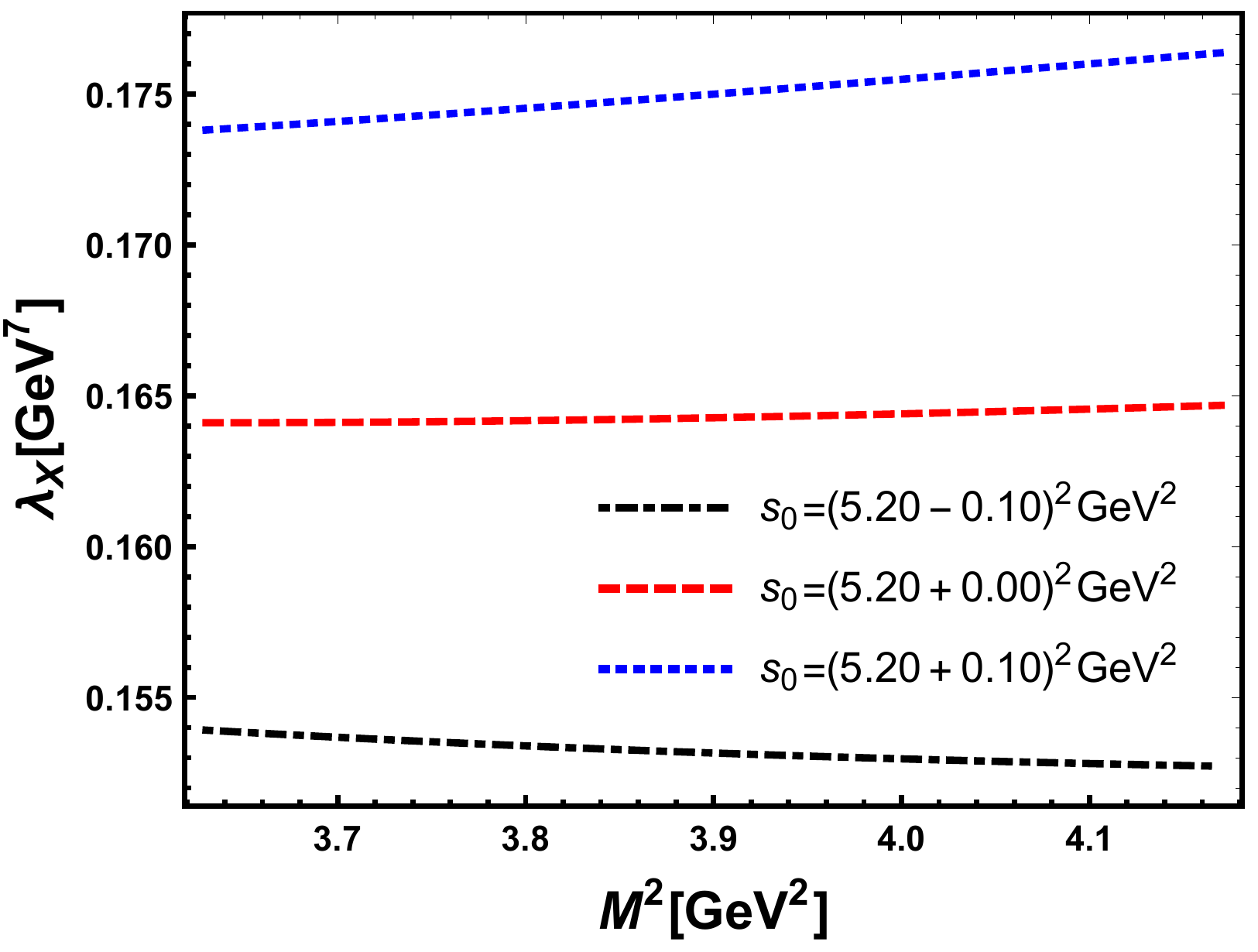}
  \caption{
  The mass [first] and the decay constant [second] of a D-wave tetraquark state $X(4700)$ as a function of the Borel parameter $M^2$ at different fixed values of $s_0$.}
\label{Fig:fig6}
\end{figure}

In Fig.\ref{Fig:fig5} and Fig.\ref{Fig:fig6} we present the results of the mass $m_X^I$ and the decay constant $f_X^I$ as functions of the parameters $M^2$ at fixed values of $s_0$$\in$$\{(5.20-0.10)^2$, $(5.20+0.0)^2$, $(5.20+0.1)^2\}$.
As we seen in the first figure in Fig.\ref{Fig:fig5} and Fig.\ref{Fig:fig6}, the yellow curves correspond to the measurements of the Belle Collaboration \cite{BESIII:2016bnd} as shown in PDG. The blue, red and black curves show clear dependence of our prediction on $s_0$ and $M^2$. By choosing appropriate parameters, our predictions are consistent with the measurements.
%the mass of $X(4700)$ provided in PDG intersect with our prediction of $m_Y$.
At a fixed point of $M^2=3.9$ $\rm GeV^2$, 
%So we could choose green line at the point $M^2=7$ represents the average result of mass.
the masses of $J_{(1)}^{X}$ and $J_{(2)}^{X}$ are
\begin{equation}
	\begin{aligned}
		\quad m_X^{(1)}=4.70^{+0.06}_{-0.05}\ \text{GeV},
	\end{aligned}
\end{equation}
and
\begin{equation}
	\begin{aligned}
		\quad  m_X^{(1)}=4.67^{+0.04}_{-0.07}\ \text{GeV},
	\end{aligned}
\end{equation}
respectively.
%Where uncertain boundaries come from black and purple line at the same point.
%Within theoretical errors $m_Y$ is in agreement with
%the measurements of the Belle Collaboration \cite{BESIII:2016bnd}.

The uncertainty comes from the various condensates, and the strange and charm quark masses.
Base on the Belle Collaboration measurements~\cite{BESIII:2016bnd}, 
$X(4700)$ has mass of $4704\pm 10^{+14}_{-24}$ MeV. 
%\deleted{We see that our mass prediction confirm our technique generalization is credible.} 
The predicted results of the two assignments are consistent with the experimental results. 
Therefore we can conclude that it might be a D-wave $cs\bar{c}\bar{s}$ tetraquark or a scalar tetraquark state.
We then extend the same technique to evaluate the decay constant of $X(4700)$,
the results of $J_{(1)}^{X}$ and $J_{(2)}^{X}$ at the same bench mark point reads
\begin{equation}
	\begin{aligned}
	\lambda_X^{(1)} \equiv m_X^{(1)}  f_X^{(1)}=0.19^{+0.013}_{-0.014}\ \text{GeV}^5,
	\end{aligned}
\end{equation}
and
\begin{equation}
	\begin{aligned}
	\lambda_X^{(2)}\equiv  m_X^{(2)} f_X^{(2)}=0.164^{+0.010}_{-0.011}\ \text{GeV}^7.
	\end{aligned}
\end{equation}
The mass and decay constant given above will be used as input parameters to find the decay width of $X$(4700)$\to J/\psi \phi$.
\subsection{The coupling constant and the decay width}
\begin{figure}[htbp]
\centering
  \includegraphics[width=7cm]{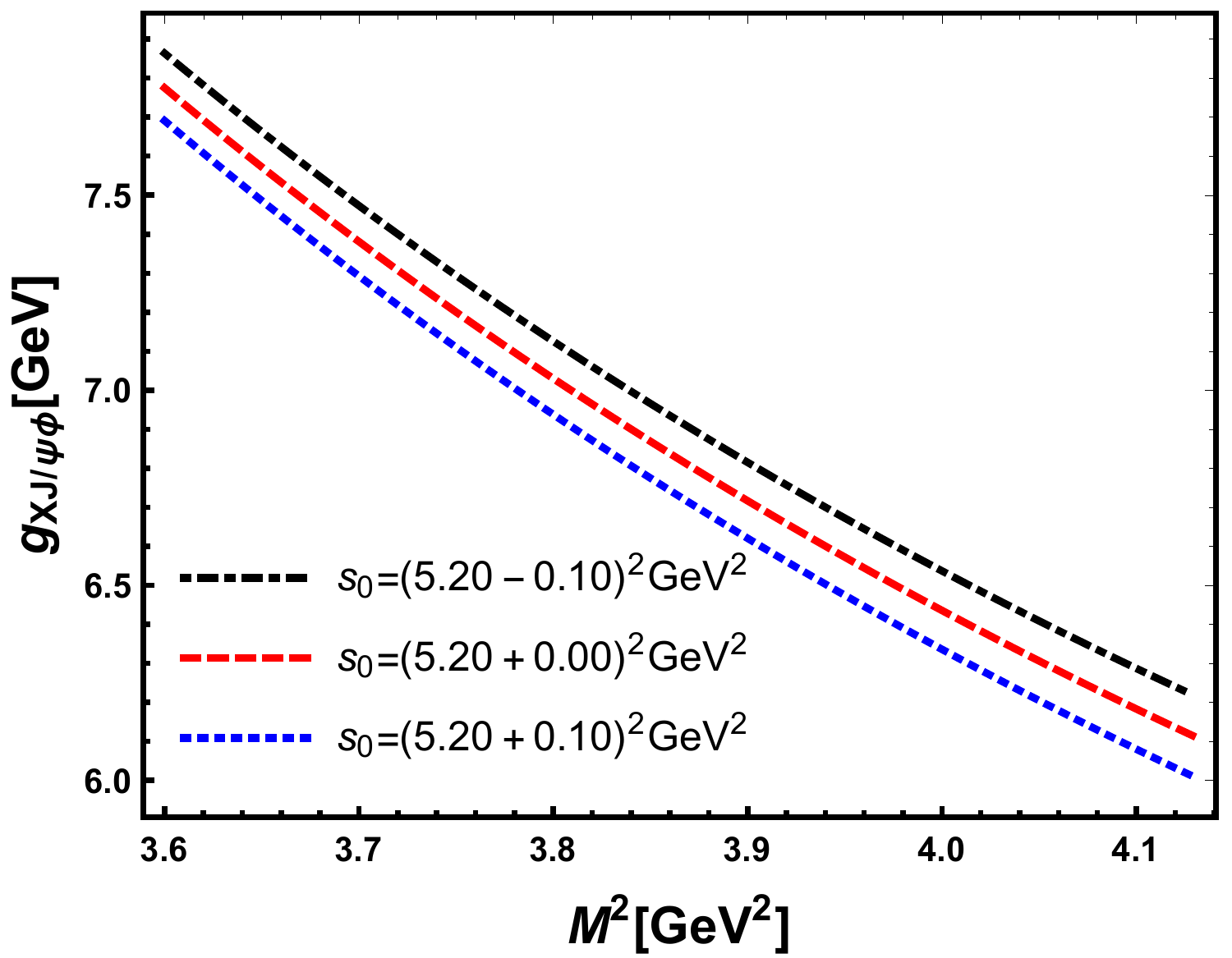}
  \caption{
  The strong coupling $g_{XJ/\psi \phi}$ of a scalar tetraquark state $X(4700)$ as a function of the Borel parameter $M^2$ at different fixed values of $s_0$.}
\label{Fig:fig7}
\end{figure}

\begin{figure}[htbp]
  \includegraphics[width=7cm]{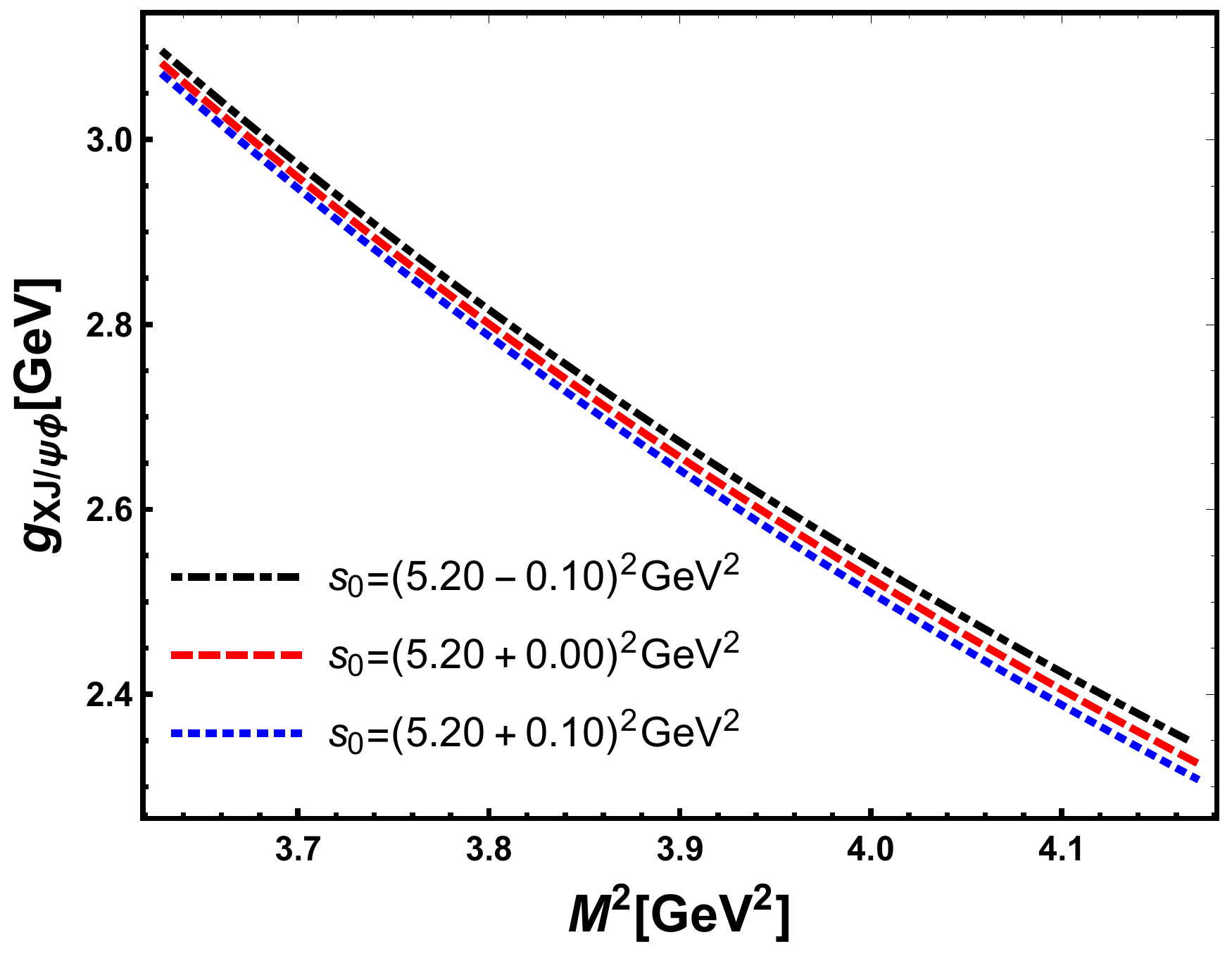}
  \caption{
  The strong coupling $g_{XJ/\psi \phi}$ of a D-wave tetraquark state $X(4700)$ as a function of the Borel parameter $M^2$ at different fixed values of $s_0$.}
\label{Fig:fig8}
\end{figure}

We start to evaluate $g_{XJ/\psi \phi}$ 
and give the decay width of $X$(4700)$\to J/\psi \phi$ in order to confirm which hypotheses are more suitable.
For $M^2$ and $s_0$, we use the same values as in the analysis of the mass. 
The result is shown in Fig.\ref{Fig:fig7} and Fig.\ref{Fig:fig8}.
Fig.\ref{Fig:fig7} provides the result for the tetraquark state.
%\deleted{The red, blue and black curves show clear dependence of our prediction on $s_0$ and $M^2$.} 
The parameters $M^2$ and $s_0$ are varied inside of the regions:$(3.60-4.13)^2\ \text{GeV}^2$ and $(5.10-5.30)^2\ \text{GeV}^2$.
%\deleted{By choosing appropriate parameters,} 
Our prediction for $g_{XJ/\psi \phi}$, take the average value, is
\begin{equation}
	\begin{aligned}
	g_{XJ/\psi \phi}=6.7^{+1.0}_{-0.8}\ \text{GeV}.
	\end{aligned}
\end{equation}
The width of this decay can be obtained by Eq.\eqref{couplingn}:
\begin{equation}
	\begin{aligned}
	\Gamma(X(4700) \rightarrow J/\psi \phi)=(109^{+35}_{-24})\ \text{MeV}.
	\end{aligned}
\end{equation}
The total width of $X$(4700) from PDG\cite{ParticleDataGroup:2020ssz} shows that
\begin{equation}
	\begin{aligned}
		\Gamma(4700)=120\pm 31^{+42}_{-33}\text{MeV},
	\end{aligned}
\end{equation}
which is close to our prediction within the error.
The result indicate that if we assign $X$(4700) as a scalar $cs\bar{c}\bar{s}$ tetraquark state $X$(4700) $\to$ $J/\psi \phi$ will the predominant process.
%\deleted{If $X$(4700) $\to$ $J/\psi \phi$ is the predominant process of $X$(4700), it shows that our calculation support the possibility that $X$(4700) could be a scalar $cs\bar{c}\bar{s}$ tetraquark state.}

For D-wave tetraquark state, the results are shown in Fig.\ref{Fig:fig8}. The prediction for $g_{XJ/\psi \phi}$ is
\begin{equation}
	\begin{aligned}
	g_{XJ/\psi \phi}=2.65^{+0.44}_{-0.33}\ \text{GeV}.
	\end{aligned}
\end{equation}
The width of its decay can be obtained by Eq.\eqref{couplingn} to be
\begin{equation}
	\begin{aligned}
	\Gamma(X(4700) \rightarrow J/\psi \phi)=(17.1^{+6.2}_{-4.0})\ \text{MeV}.
	\end{aligned}
\end{equation}
which is much smaller than the width of $X$(4700) in PDG. Our results illustrates that,
in this case, if one assign $X$(4700) as a D-wave tetraquark state,
$X$(4700) $\to$ $J/\psi \phi$ will be the minor decay channel.
Therefore the result imply that there exist multiple decays that not yet found in experiments.
%\deleted{but one of some multiples when we assign D-wave tetraquark state for $X$(4700).}

%\deleted{In conclusion, assigning scalar $cs\bar{c}\bar{s}$ tetraquark state for $X$(4700)
%will lead to result that there is only one dominant decay process for $X$(4700),
%and assigning D-wave tetraquark state for $X$(4700) will imply that there exist multiple decays that not yet found in experiments.
%We expect the future experiments will provide more information.}

\section{Summary}
\label{sec:summary}

In this work we assign $X$(4700) as a D-wave tetraquark state and a scalar tetraquark state 
to study the mass and the decay constant of $X(4700)$, and also it's decay $X(4700)\to J/\psi\phi$. 
The mass of $X(4700)$ is evaluated through two-point sum rules,
and both the results of the two assignments are 
in agreement with the mass of $X(4700)$ in PDG. 
We also calculate the decay constant of $X(4700)$ in the SVZ sum rules.
We then perform the calculation of the coupling constant $g_{X J/\psi\phi}$
and the decay width by using the approach of the light-cone sum rules.
We find that being a scalar tetraquark state will make $X(4700)$ most likely decay into $J/\psi\phi$
since in this case $X(4700)\to J/\psi\phi$ is the predominant decay process, while being a D-wave tetraquark state,
the decay of $X(4700)\to J/\psi\phi$ is much smaller and become a non-significant decay channel.

\begin{acknowledgments}
Hao Sun is supported by the National Natural Science Foundation of China (Grant No.12075043, No.12147205).
\end{acknowledgments}

\section{Appendix}

\subsection{The relations between the light-cone distribution amplitudes (LCDAs) and the matrix elements}
\label{appendix:A}

The matrix elements of the $\phi$ can be expanded in terms of the corresponding distribution amplitudes. Here we provide the expressions for the $\left\langle\phi(P)|\bar{q}(0)\Gamma^aq(0)|0\right\rangle$ type matrix elements \cite{Ball:1996tb}:
\begin{eqnarray}
&&\left\langle \phi(P, \lambda)\left|\bar{q}(0) \gamma_{\mu} q(0)\right| 0\right\rangle = f_{\phi}^{\|} m_{\phi} \epsilon_{\mu}^{*(\lambda)} \\
&&\left\langle \phi(P, \lambda)\left|\bar{q}(0) \sigma_{\mu \nu} q(0)\right| 0\right\rangle=i f_{\phi}^{\perp}\left(e_{\mu}^{*(\lambda)} P_{\nu}-e_{\nu}^{*(\lambda)} P_{\mu}\right). \ \ \ \ \ \ \ \
\end{eqnarray}
We also provide the expressions for $\langle \phi(P)|\bar{q}(0)gG_{\alpha \beta}\Gamma^a q(0)|0 \rangle$ type matrix elements \cite{Ball:2007zt,Ball:2007rt}:
\begin{eqnarray} 
&&\left\langle \phi(P, \lambda)\left|\bar{q}(0) g \widetilde{G}_{\alpha \beta} \gamma_{\mu} \gamma_{5} q(0)\right| 0\right\rangle\\ \nonumber
&&= f_{\phi}^{\|} m_{\phi} \zeta_{3 \phi}^{\|}[e_{\alpha}^{(\lambda)}\left(P_{\beta} P_{\mu}-\frac{1}{3} m_{\phi}^{2} g_{\beta \mu}\right)-(\alpha \leftrightarrow \beta)] \\ \nonumber
&&+\frac{1}{3} f_{\phi}^{\|} m_{\phi}^{3} \zeta_{4 \phi}^{\|}\left(e_{\alpha}^{(\lambda)} g_{\beta \mu}-e_{\beta}^{(\lambda)} g_{\alpha \mu}\right),\\ 
&&\left\langle \phi(P, \lambda)\left|\bar{q}(0) g G_{\alpha \beta} i \gamma_{\mu} q(0)\right| 0\right\rangle\\ \nonumber
&&=-f_{\phi}^{\|} m_{\phi} \kappa_{3 \phi}^{\|}[e_{\alpha}^{(\lambda)}\left(P_{\beta} P_{\mu}-\frac{1}{3} m_{\phi}^{2} g_{\beta \mu}\right)-(\alpha \leftrightarrow \beta)] \\ \nonumber
&&-\frac{1}{3} f_{\phi}^{\|} m_{\phi}^{3} \kappa_{4 \phi}^{\|}\left(e_{\alpha}^{(\lambda)} g_{\beta \mu}-e_{\beta}^{(\lambda)} g_{\alpha \mu}\right), \\ 
&&\left\langle \phi(P, \lambda)\left|\bar{q}(0) g G_{\alpha \beta} q(0)\right| \right\rangle\\ \nonumber
&&=-i f_{\phi}^{\perp} m_{\phi}^{2} \zeta_{4 \phi}^{\perp}\left(\epsilon_{\alpha}^{(\lambda)} P_{\beta}-\epsilon_{\beta}^{(\lambda)} P_{\alpha}\right),\\ 
&&\left\langle \phi(P, \lambda)\left|\bar{q}(0) g \widetilde{G}_{\alpha \beta} \gamma_{5} q(0)\right| 0\right\rangle\\ \nonumber
&&=f_{\phi}^{\perp} m_{\phi}^{2} \widetilde{\zeta}_{4 \phi}^{\perp}\left(\epsilon_{\alpha}^{(\lambda)} P_{\beta}-\epsilon_{\beta}^{(\lambda)} P_{\alpha}\right), \\ 
&&\left\langle \phi(P, \lambda)\left|\bar{q}(0) g G_{\alpha \mu} \sigma_{\beta \mu} q(0)\right| \right\rangle\\ \nonumber
&&=f_{\phi}^{\perp} m_{\phi}^{2}[\frac{1}{2} \kappa_{3 \phi}^{\perp}\left(\epsilon_{\alpha}^{(\lambda)} P_{\beta}+\epsilon_{\beta}^{(\lambda)} P_{\alpha}\right) \\ \nonumber
&&+\kappa_{4 \phi}^{\perp}\left(\epsilon_{\alpha}^{(\lambda)} P_{\beta}-\epsilon_{\beta}^{(\lambda)} P_{\alpha}\right)],
\end{eqnarray}
where the dual gluon field strength tensor is defined as $\widetilde{G}_{\mu \nu}=\frac{1}{2} \epsilon_{\mu \nu \rho \sigma} G^{\rho \sigma}$. The generic notations of $\zeta$ are G-conserving and $\kappa$ are G-breaking parameters.
$\zeta_{3\phi}$, $\widetilde{\zeta}_{3\phi}$, $\kappa_{3\phi}$ are twist-3 and $\zeta_{4\phi}$, $\widetilde{\zeta}_{4\phi}$, $\kappa_{4\phi}$ are twist-4 parameters given in \cite{Ball:2007zt}.

The covariant derivative is defined as
\begin{equation}
\begin{aligned}
\overleftarrow{D}_\mu=\partial_\mu +igT^aA_\mu^a,
\end{aligned}
\end{equation}
and it is also valid in the Fock-Schwinger gauge
\begin{equation}
	\begin{aligned}
		x^\mu A_\mu^a(x)=0,
	\end{aligned}
\end{equation}
after performing some deduction, we can derive the following relations \cite{Gubler:2013moa}:
\begin{equation}
\begin{aligned}
A_{\mu}^{a}(x)=& \frac{1}{2} x^{v} G_{v \mu}^{a}(0)+\frac{1}{3} x^{v} x^{\alpha}\left[D_{\alpha} G_{v \mu}(0)\right]^{a} \\
&+\frac{1}{8} x^{v} x^{\alpha} x^{\beta}\left[D_{\alpha} D_{\beta} G_{v \mu}(0)\right]^{a}+\cdots.
\end{aligned}
\end{equation}
Therefore, insert the gluonic fields back to the covariant derivative and calculate  $\braket{\phi(P)|\bar{q}(0)\overleftarrow{D}_\mu\overleftarrow{D}_\nu\Gamma^aq(0)|0}$, we obtain the desired results:
\begin{eqnarray}
		&&\braket{\phi(P)|\bar{q}(0)\overleftarrow{D}_\mu\overleftarrow{D}_\nu\Gamma^aq(0)|0}\\ \nonumber
		&&=\braket{\phi(P)|\partial_\mu\partial_\nu[\bar{q}(x)]|_{x=0}\Gamma^aq(0)|0}\\ \nonumber
		&&+\braket{\phi(P)|\bar{q}(0)\frac{igT^b}{2}G^b_{\mu\nu}\Gamma^aq(0)|0}\\ \nonumber
		&&=\partial_\mu\partial_\nu[\braket{\phi(P)|\bar{q}(x)\Gamma^aq(0)|0}]|_{x=0}\\ \nonumber
		&&+\frac{i}{2}\braket{\phi(P)|\bar{q}(0)gG_{\mu\nu}\Gamma^aq(0)|0},
\end{eqnarray}
where $\braket{\phi(P)|\bar{q}(0)gG_{\mu\nu}\Gamma^aq(0)|0}$ type matrix elements are given before. In addition, $\braket{\phi(P)|\bar{q}(x)\Gamma^aq(0)|0}$ type expressions are provided below \cite{Ball:1996tb}:
\begin{eqnarray}
&& \left\langle \phi(P, \lambda)\left|\bar{q}(x) \sigma_{\rho \nu} q(0)\right| 0\right\rangle \\ \nonumber
&&=-i\left(e_{\rho}^{(\lambda)} P_{\nu}-e_{\nu}^{(\lambda)} P_{\rho}\right) \times f_{\phi}^{\perp} \int_{0}^{1} d u e^{i u p x} \phi_{2}^\perp(u,\mu),\\ 
&&\left\langle \phi(P, \lambda)\left|\bar{q}(x) \gamma_{\theta} \gamma_{5} q(0)\right| 0\right\rangle \\ \nonumber
&&=-\frac{1}{4} \epsilon_{\theta \nu \rho \sigma} e^{(\lambda) \nu} P^{\rho} x^{\sigma} f_{\phi} m_{\phi} \times \int_{0}^{1} d u e^{i u p x} g_{\perp}^{(a)}(u, \mu), \\
&&\left\langle \phi(P, \lambda)\left|\bar{q}(x) \gamma_{\rho} q(0)\right| 0\right\rangle \\ \nonumber
&&= P_{\rho}\left(e^{(\lambda)} x\right) f_{\phi} m_{\phi} \times \int_{0}^{1} d u e^{i u p x} \Phi_{\|}(u, \mu) \\ \nonumber
&&+e_{\rho}^{(\lambda)} f_{\phi} m_{\phi} \int_{0}^{1} d u e^{i u p x} g_{\perp}^{(v)}(u, \mu),
\end{eqnarray}
where
\begin{eqnarray}
&& g_{\perp}^{(v), \text { twist } 2}(u, \mu) \\ \nonumber
&&=\frac{1}{2}\left[\int_{0}^{u} d y \frac{\phi^{\|}_2(y, \mu)}{\bar{y}}+\int_{u}^{1} d y \frac{\phi^{\|}_2(y, \mu)}{y}\right], \\
&& 
g_{\perp}^{(a), \text { twist } 2}(u, \mu) \\ \nonumber
&&=2\left[\bar{u} \int_{0}^{u} d y \frac{\phi^{\|}_2(y, \mu)}{\bar{y}}+u \int_{u}^{1} d y \frac{\phi^{\|}_2(y, \mu)}{y}\right],
\end{eqnarray}
and
\begin{eqnarray}
&&
\Phi_{\|}(u, \mu) \\ \nonumber
&&=\frac{1}{2}\left[\bar{u} \int_{0}^{u} d y \frac{\phi^{\|}_2(y, \mu)}{\bar{y}}-u \int_{u}^{1} d y \frac{\phi^{\|}_2(y, \mu)}{y}\right].
\end{eqnarray}
Here $\bar{u}=1-u$, $\bar{y}=1-y$. In term of Gegenbauer polynomials, $\phi_{2}^{\|, \perp}$ is given as \cite{Ball:2007zt}
\begin{equation}
\phi_{2}^{\|, \perp}(u, \mu)=6 u \bar{u}\left\{1+\sum_{n=1}^{\infty} a_{n}^{\|, \perp}(\mu) C_{n}^{3 / 2}(2 u-1)\right\},
\end{equation}
the Gegenbauer polynomials $C_{n}^{3 / 2}(x)$ and coefficients $a_n$ at the renormalisation scale $\mu$ are given in details in \cite{Ball:1996tb}.

\subsection{Spectral densities}
\label{appendix:B}

In this section we provide the spectral densities for $J_{(1)}$ and $J_{(2)}$.
In the following expressions, $\mathrm{H}(x)$ is defined as:
\begin{eqnarray}
    \mathrm{H}(x)=
    \begin{cases}
    0&x\ge0\\
    1&x< 0 .
    \end{cases}
\end{eqnarray}
The spectral density for $J_{(1)}$ can be divide into:
\begin{eqnarray} \nonumber
\Tilde{\rho}_{(1)}^{OPE}(\hat{s})&=&
\Tilde{\rho}_{(1)}^{pert}(\hat{s})+\Tilde{\rho}_{(1)}^{\braket{\bar{q}q}}(\hat{s})
+\Tilde{\rho}_{(1)}^{\braket{\frac{\alpha_s}{\pi}GG}}(\hat{s})\\
&+&\tilde{\rho}_{(1)}^{\braket{\bar{q}g_s\sigma Gq}}(\hat{s})+\tilde{\rho}_{(1)}^{\braket{\bar{q}q}^2}(\hat{s})+\Tilde{\rho}_{(1)}^{\braket{\frac{\alpha_s}{\pi}GG} \braket{\bar{q}q}}(\hat{s}) \ \ ~~~~
\end{eqnarray}
with
\begin{eqnarray} \nonumber
&&\Tilde{\rho}_{(1)}^{pert}(\hat{s})=
\int_0^1dx\int_0^1dy\frac{y \left(m_c^2+\hat{s} x (x (-y)+x+y-1)\right)^2}{768 \pi ^6 (x-1)^3 x^3 (y-1)^3}\\ \nonumber
&\times& 
\left(
m^4 y \left(3 y^3+7 y^2-3 y+3\right)-12 m_c^3 m_s y \left(y^2-1\right)\right. \\ \nonumber
&-&2 m_c^2 (x - 1) x (y - 1) (36 m_s^2 (y - 1)  \\ \nonumber
&+&  \hat{s} y (9 y^3 + 20 y^2 + 9 y + 6))\\ \nonumber
&+&30 m_c m_s \hat{s} (x-1) x (y-1)^2 y (y+1)\\ \nonumber
&+& \left. 3 \hat{s}^2 (x - 1)^2 x^2 (y - 1)^2 y (7 y^3 + 15 y^2 + 9 y + 3)\right)\\
&\times& \mathrm{H}\left(y \left(x \hat{s} (x (-y)+x+y-1)+m_c^2\right)\right),
\end{eqnarray}
\begin{eqnarray} \nonumber
&&\Tilde{\rho}_{(1)}^{\braket{\bar{q}q}}(\hat{s})=-\int_0^1dx\int_0^1dy\frac{\braket{\bar{q}q}}{16 \pi ^4 (x-1)^2 x^2 (y-1)}\\ \nonumber
&\times& \left(2 m^5 y (y+1) \right. \\ \nonumber
&-& 2 m^4 \text{mq} (x-1) x \left(3 y^3+5 y^2-3 y+5\right) \\ \nonumber
&+&2 m_c^3 (x-1) x \left(y^2-1\right) \left(m_s^2-3 \hat{s} y\right)\\ \nonumber
&+&2 m_c^2 m_s \hat{s} (x-1)^2 x^2 \left(12 y^4+9 y^3-15 y^2+y-7\right)\\  \nonumber
&+&m \hat{s} (x-1)^2 x^2 (y-1)^2 (y+1) \left(4 \hat{s} y-3 m_q^2\right)\\ \nonumber
&-&\left. 4 m_s \hat{s}^2 (x-1)^3 x^3 (y-1)^2 \left(5 y^3+9 y^2+5 y+1\right)\right)\\
&\times & \mathrm{H}\left(y \left(x \hat{s} (x (-y)+x+y-1)+m_c^2\right)\right) ,
\end{eqnarray}
\begin{eqnarray} \nonumber
&&\Tilde{\rho}_{(1)}^{\braket{\frac{\alpha_s}{\pi}GG}}(\hat{s})=
\int_0^1dx\int_0^1dy\frac{\braket{\frac{\alpha_s}{\pi}GG}}{512 \pi ^4 (x-1) x (y-1)}\\ \nonumber
&&\left(\times m_c^4 \left(3 y^3+7 y^2+y+1\right)\right.\\ \nonumber
&-&m^2 \hat{s} (x-1) x \left(12 y^4+15 y^3-17 y^2-7 y-3\right)\\ \nonumber
&+&\left. 2 \hat{s}^2 (x-1)^2 x^2 (y-1)^2 \left(5 y^3+11 y^2+5 y+1\right)\right)\\ \nonumber
&\times& \mathrm{H}\left(y \left(x \hat{s} (x (-y)+x+y-1)+m_c^2\right)\right) ,
\end{eqnarray}
\begin{eqnarray} \nonumber
&&\Tilde{\rho}_{(1)}^{\braket{\bar{q}g_s\sigma Gq}}(\hat{s})=
\int_0^1dx\int_0^1dy\frac{\braket{\bar{q}g_s\sigma Gq}}{192 \pi ^4 (x-1) x}\\ \nonumber
&\times& \left(2 m_c^3 (2 x-7) (y+1) \right. \\ \nonumber
&+& 8 m_c^2 m_s (x-1) x \left(3 y^2+4 y+3\right)\\ \nonumber
&-& 3 m_c \hat{s} \left(2 x^2-9 x+7\right) x \left(y^2-1\right)\\ \nonumber
&-&\left. 24 m_s \hat{s} (x-1)^2 x^2 \left(2 y^3+y^2-y-2\right)\right)\\ \nonumber
&\times& \mathrm{H}\left(y \left(x \hat{s} (x (-y)+x+y-1)+m_c^2\right)\right)\\ \nonumber
&+&\int_0^1dx\left(\frac{ \braket{\bar{q}g_s\sigma Gq} \left(64 m_c^2 m_s-8 m_s \hat{s} x (x+3)\right)}{384 \pi ^4}\right.\\ \nonumber
&+&\left.\frac{ \braket{\bar{q}g_s\sigma Gq} \left(m \left(-2 m_0 x+m_0+8 m_s^2\right)\right)}{384 \pi ^4}\right)\\
&\times& \mathrm{H}(m^2-x (x+1) s) ,
\end{eqnarray}
\begin{eqnarray} \nonumber
&&\Tilde{\rho}_{(1)}^{\braket{\bar{q}q}^2}(\hat{s})=\int_0^1dx\int_0^1dy \\ \nonumber
&&\mathrm{H}\left(y \left(x \hat{s} (x (-y)+x+y-1)+m_c^2\right)\right)\\ \nonumber
&\times& (-\frac{ g_s^2 \braket{\bar{q}q}^2 \left(m_c^2 \left(3 y^2+4 y+3\right)\right)}{162 \pi^4}\\ \nonumber
&-&\frac{ g_s^2 \braket{\bar{q}q}^2 \left(-3 \hat{s} (x-1) x \left(2 y^3+y^2-y-2\right)\right)}{162 \pi^4})\\ \nonumber
&+&\int_0^1dx\frac{ \braket{\bar{q}q}^2 }{324 \pi ^4}\mathrm{H}(m^2-x (x+1) s)\\ \nonumber
&\times& \left(g_s^2 \left(-2 m_c^2-m_c m_s+\hat{s} (x+3) x\right)\right.\\
&+&\left. 27 \pi ^2 \left(4 m_c^2+2 m_c m_s-m_s^2 (x-6) x\right)\right) ,
\end{eqnarray}
\begin{eqnarray} \nonumber
&&\tilde{\rho}_{(1)}^{\braket{\frac{\alpha_s}{\pi}GG} \braket{\bar{q}q}}(\hat{s})=\int_0^1dx\frac{im_c  \braket{\frac{\alpha_s}{\pi}GG}\braket{\bar{q}q}}{144 \pi ^3} \\
&\mathrm{H} & (m^2-x (x+1) s) .
\end{eqnarray}
The spectral density for $J_{(2)}$ can be divide into:
\begin{eqnarray} \nonumber
\Tilde{\rho}^{OPE}_{(2)}(\hat{s})&=&  \nonumber
\Tilde{\rho}_{(2)}^{\braket{\frac{\alpha_s}{\pi}GG}}(\hat{s})+\Tilde{\rho}_{(2)}^{\braket{\bar{q}g_s\sigma Gq}}(\hat{s}) \\ \nonumber
&+&
\Tilde{\rho}_{(2)}^{\braket{\frac{\alpha_s}{\pi}GG}\braket{\bar{q}q}}(\hat{s})+\Tilde{\rho}_{(2)}^{\braket{\bar{q}g_s\sigma Gq}\braket{\bar{q}q}}(\hat{s}) \\
&+&\tilde{\rho}_{2}^{\braket{\bar{q}g_s\sigma Gq}\braket{\frac{\alpha_s}{\pi}GG}}(\hat{s})+\Tilde{\rho}_{(2)}^{\braket{\bar{q}q}^2\braket{\frac{\alpha_s}{\pi}GG}}(\hat{s})
\end{eqnarray}
with
\begin{eqnarray} \nonumber
&&\Tilde{\rho}_{(2)}^{\braket{\frac{\alpha_s}{\pi}GG}}(\hat{s})=\int_0^1\int_0^1dxdy\frac{i \braket{\frac{\alpha_s}{\pi}GG} y^2}{9216 \pi ^4 (x-1)^4 x^3 (y-1)^3}\\ \nonumber
&\times& \left(m^8 \left(2 x \left(18 y^4+54 y^3+49 y^2+33 y-3\right)\right.\right.\\ \nonumber
&+&\left.36 y^4+117 y^3+109 y^2+34 y+6\right)\\ \nonumber
&-&12 m_c^6 \hat{s} x (y-1) \left(x^2 \left(30 y^4+84 y^3+79 y^2+38 y-3\right)\right.\\ \nonumber
&+& 2 x \left(3 y^3+2 y^2-7 y+2\right) \\ \nonumber
&-&\left. 30 y^4-90 y^3-83 y^2-24 y-1\right)\\ \nonumber
&+&12 m_c^4 \hat{s}^2 (x-1)^2 x^2 (y-1)^2\\ \nonumber
&\times& \left(x \left(90 y^4+240 y^3+224 y^2+90 y-6\right)\right.\\ \nonumber
&+&\left. \left(90 y^3+255 y^2+229 y+64\right) y\right)\\ \nonumber
&-&4 m_c^2 \hat{s}^3 (x-1)^3 x^3 (y-1)^3\\ \nonumber
&\times&\left(x \left(315 y^4+810 y^3+740 y^2+264 y-15\right)\right.\\ \nonumber
&+&\left. 315 y^4+855 y^3+745 y^2+202 y-3\right)\\ \nonumber
&+&3 \hat{s}^4 (x-1)^4 x^4 (y-1)^4\\ \nonumber
&\times& \left(2 x \left(84 y^4+210 y^3+187 y^2+61 y-3\right)\right.\\ \nonumber
&+&\left.\left.+168 y^4+441 y^3+373 y^2+98 y-2\right)\right)\\	
&\times& \mathrm{H}\left(y \left(x \hat{s} (x (-y)+x+y-1)+m_c^2\right)\right),
\end{eqnarray}
\begin{eqnarray} \nonumber
&&\Tilde{\rho}_{(2)}^{\braket{\bar{q}g_s\sigma Gq}}(\hat{s})=\int_0^1\int_0^1dxdy\frac{5 m_s\braket{\bar{q}g_s\sigma Gq}}{576 \pi ^4 (x-1)^2 x^2 (y-1)^2}\\ \nonumber
&\times& 2 \left(
m_c^6 \left(6 y^3+7 y^2+8 y-1\right) y\right.\\ \nonumber
&-& m_c^4 (x-1) x (y-1) \left(3 s y \left(24 y^3+33 y^2+26 y-3\right)\right.\\ \nonumber
&-&\left. m_0 \left(12 y^3+15 y^2+14 y-1\right)\right)\\ \nonumber
&+&m_c^2 \hat{s} (x-1)^2 x^2 (y-1)^2 \left(12 \hat{s} y \left(10 y^3+15 y^2+10 y-1\right)\right.\\ \nonumber
&-&\left. m_0 \left(48 y^3+69 y^2+50 y-3\right)\right)\\ \nonumber
&+&\hat{s}^2 (x-1)^3 x^3 (y-1)^3 \left(m_0 \left(40 y^3+62 y^2+40 y-2\right)\right.\\ \nonumber
&-&\left.\left.\hat{s} y \left(60 y^3+95 y^2+58 y-5\right)\right)\right)\\
&\times& \mathrm{H}\left(y \left(x \hat{s} (x (-y)+x+y-1)+m_c^2\right)\right),
\end{eqnarray}
\begin{eqnarray} \nonumber
&&\tilde{\rho}_{(2)}^{\braket{\frac{\alpha_s}{\pi}GG}\braket{\bar{q}q}}(\hat{s})\\ \nonumber
&=&\int_0^1\int_0^1dxdy \mathrm{H}\left(y \left(x \hat{s} (x (-y)+x+y-1)+m_c^2\right)\right)\\ \nonumber
&\times& \frac{-m_c\braket{\frac{\alpha_s}{\pi}GG}\braket{\bar{q}q} \left(m_c^2+\hat{s} x (x (-y)+x+y-1)\right)}{16 \pi ^2 (x-1)^2 x (y-1)}\\ \nonumber
&\times& \left(m_c^2 (y+1) y+2 m_c m_s (x-1) (y-1) \right. \\
&-& \left. 2 \hat{s} (x-1) x y \left(y^2-1\right)\right),
\end{eqnarray}
\begin{eqnarray} \nonumber
&&\Tilde{\rho}_{(2)}^{\braket{\bar{q}g_s\sigma Gq}\braket{\bar{q}q}}(\hat{s})\\ \nonumber
&=&\int_0^1\int_0^1dxdy \mathrm{H}\left(y \left(x \hat{s} (x (-y)+x+y-1)+m_c^2\right)\right)\\ \nonumber
&\times& \frac{-5 g_s^2 \braket{\bar{q}g_s\sigma Gq} \braket{\bar{q}q}}{3888 \pi ^4 (x-1) x (y-1)}\\ \nonumber
&\times& \left(m_c^4 \left(12 y^3+15 y^2+14 y-1\right)\right.\\ \nonumber
&-&m_c^2 \hat{s} (x-1) x \left(48 y^4+21 y^3-19 y^2-53 y+3\right)\\
&+&\left.2 \hat{s}^2 (x-1)^2 x^2 (y-1)^2 \left(20 y^3+31 y^2+20 y-1\right)\right), \ \ \ \ \ ~~~
\end{eqnarray}
\begin{eqnarray} \nonumber
&&\Tilde{\rho}_{2}^{\braket{\bar{q}g_s\sigma Gq}\braket{\frac{\alpha_s}{\pi}GG}}(\hat{s})\\ \nonumber
&=&\int_0^1\int_0^1dxdy \mathrm{H}\left(y \left(x \hat{s} (x (-y)+x+y-1)+m_c^2\right)\right)\\ \nonumber
&\times&\left(
-\frac{5 \braket{\frac{\alpha_s}{\pi}GG} m_c \braket{\bar{q}g_s\sigma Gq}(y+1) \left(2 m^2\right)}{768 \pi ^2 (x-1)}\right.\\ \nonumber
&-&  \frac{5 \braket{\frac{\alpha_s}{\pi}GG} m_c \braket{\bar{q}g_s\sigma Gq}}{768 \pi ^2 (x-1)}  \\
&\times& \left. (y+1) \left(3 \hat{s} x (x (-y)+x+y-1)\right) \right),
\end{eqnarray}
\begin{eqnarray} \nonumber
&&\Tilde{\rho}_{(2)}^{\braket{\bar{q}q}^2\braket{\frac{\alpha_s}{\pi}GG}}(\hat{s})\\ \nonumber
&=&\int_0^1dx \mathrm{H}\left(m_c^2-x (x+1) \hat{s}\right)\\
&&\times\frac{1}{6} \braket{\frac{\alpha_s}{\pi}GG} m_c^2 \braket{\bar{q}q}^2-\frac{1}{12}  \braket{\frac{\alpha_s}{\pi}GG} m_c m_s\braket{\bar{q}q}^2 x . ~~~~~~~~
\end{eqnarray}

\bibliography{ref}

\begin{thebibliography}{50}%
\makeatletter
\providecommand \@ifxundefined [1]{%
 \@ifx{#1\undefined}
}%
\providecommand \@ifnum [1]{%
 \ifnum #1\expandafter \@firstoftwo
 \else \expandafter \@secondoftwo
 \fi
}%
\providecommand \@ifx [1]{%
 \ifx #1\expandafter \@firstoftwo
 \else \expandafter \@secondoftwo
 \fi
}%
\providecommand \natexlab [1]{#1}%
\providecommand \enquote  [1]{``#1''}%
\providecommand \bibnamefont  [1]{#1}%
\providecommand \bibfnamefont [1]{#1}%
\providecommand \citenamefont [1]{#1}%
\providecommand \href@noop [0]{\@secondoftwo}%
\providecommand \href [0]{\begingroup \@sanitize@url \@href}%
\providecommand \@href[1]{\@@startlink{#1}\@@href}%
\providecommand \@@href[1]{\endgroup#1\@@endlink}%
\providecommand \@sanitize@url [0]{\catcode `\\12\catcode `\$12\catcode
  `\&12\catcode `\#12\catcode `\^12\catcode `\_12\catcode `\%12\relax}%
\providecommand \@@startlink[1]{}%
\providecommand \@@endlink[0]{}%
\providecommand \url  [0]{\begingroup\@sanitize@url \@url }%
\providecommand \@url [1]{\endgroup\@href {#1}{\urlprefix }}%
\providecommand \urlprefix  [0]{URL }%
\providecommand \Eprint [0]{\href }%
\providecommand \doibase [0]{http://dx.doi.org/}%
\providecommand \selectlanguage [0]{\@gobble}%
\providecommand \bibinfo  [0]{\@secondoftwo}%
\providecommand \bibfield  [0]{\@secondoftwo}%
\providecommand \translation [1]{[#1]}%
\providecommand \BibitemOpen [0]{}%
\providecommand \bibitemStop [0]{}%
\providecommand \bibitemNoStop [0]{.\EOS\space}%
\providecommand \EOS [0]{\spacefactor3000\relax}%
\providecommand \BibitemShut  [1]{\csname bibitem#1\endcsname}%
\let\auto@bib@innerbib\@empty
%</preamble>
\bibitem [{\citenamefont {Aaij}\ \emph
  {et~al.}(2017{\natexlab{a}})\citenamefont {Aaij} \emph
  {et~al.}}]{LHCb:2016axx}%
  \BibitemOpen
  \bibfield  {author} {\bibinfo {author} {\bibfnamefont {R.}~\bibnamefont
  {Aaij}} \emph {et~al.} (\bibinfo {collaboration} {LHCb}),\ }\href {\doibase
  10.1103/PhysRevLett.118.022003} {\bibfield  {journal} {\bibinfo  {journal}
  {Phys. Rev. Lett.}\ }\textbf {\bibinfo {volume} {118}},\ \bibinfo {pages}
  {022003} (\bibinfo {year} {2017}{\natexlab{a}})},\ \Eprint
  {http://arxiv.org/abs/1606.07895} {arXiv:1606.07895 [hep-ex]} \BibitemShut
  {NoStop}%
\bibitem [{\citenamefont {Aaij}\ \emph
  {et~al.}(2017{\natexlab{b}})\citenamefont {Aaij} \emph
  {et~al.}}]{LHCb:2016nsl}%
  \BibitemOpen
  \bibfield  {author} {\bibinfo {author} {\bibfnamefont {R.}~\bibnamefont
  {Aaij}} \emph {et~al.} (\bibinfo {collaboration} {LHCb}),\ }\href {\doibase
  10.1103/PhysRevD.95.012002} {\bibfield  {journal} {\bibinfo  {journal} {Phys.
  Rev. D}\ }\textbf {\bibinfo {volume} {95}},\ \bibinfo {pages} {012002}
  (\bibinfo {year} {2017}{\natexlab{b}})},\ \Eprint
  {http://arxiv.org/abs/1606.07898} {arXiv:1606.07898 [hep-ex]} \BibitemShut
  {NoStop}%
\bibitem [{\citenamefont {Gui}\ \emph {et~al.}(2018)\citenamefont {Gui},
  \citenamefont {Lu}, \citenamefont {L\"u}, \citenamefont {Zhong},\ and\
  \citenamefont {Zhao}}]{Gui:2018rvv}%
  \BibitemOpen
  \bibfield  {author} {\bibinfo {author} {\bibfnamefont {L.-C.}\ \bibnamefont
  {Gui}}, \bibinfo {author} {\bibfnamefont {L.-S.}\ \bibnamefont {Lu}},
  \bibinfo {author} {\bibfnamefont {Q.-F.}\ \bibnamefont {L\"u}}, \bibinfo
  {author} {\bibfnamefont {X.-H.}\ \bibnamefont {Zhong}}, \ and\ \bibinfo
  {author} {\bibfnamefont {Q.}~\bibnamefont {Zhao}},\ }\href {\doibase
  10.1103/PhysRevD.98.016010} {\bibfield  {journal} {\bibinfo  {journal} {Phys.
  Rev. D}\ }\textbf {\bibinfo {volume} {98}},\ \bibinfo {pages} {016010}
  (\bibinfo {year} {2018})},\ \Eprint {http://arxiv.org/abs/1801.08791}
  {arXiv:1801.08791 [hep-ph]} \BibitemShut {NoStop}%
\bibitem [{\citenamefont {Fern\'andez}\ \emph {et~al.}(2016)\citenamefont
  {Fern\'andez}, \citenamefont {Entem}, \citenamefont {Ortega},\ and\
  \citenamefont {Segovia}}]{Fernandez:2016bqr}%
  \BibitemOpen
  \bibfield  {author} {\bibinfo {author} {\bibfnamefont {F.}~\bibnamefont
  {Fern\'andez}}, \bibinfo {author} {\bibfnamefont {D.~R.}\ \bibnamefont
  {Entem}}, \bibinfo {author} {\bibfnamefont {P.~G.}\ \bibnamefont {Ortega}}, \
  and\ \bibinfo {author} {\bibfnamefont {J.}~\bibnamefont {Segovia}},\ }\href
  {\doibase 10.22323/1.289.0054} {\bibfield  {journal} {\bibinfo  {journal}
  {PoS}\ }\textbf {\bibinfo {volume} {CHARM2016}},\ \bibinfo {pages} {054}
  (\bibinfo {year} {2016})},\ \Eprint {http://arxiv.org/abs/1611.08534}
  {arXiv:1611.08534 [hep-ph]} \BibitemShut {NoStop}%
\bibitem [{\citenamefont {Ortega}\ \emph {et~al.}(2016)\citenamefont {Ortega},
  \citenamefont {Segovia}, \citenamefont {Entem},\ and\ \citenamefont
  {Fern\'andez}}]{Ortega:2016hde}%
  \BibitemOpen
  \bibfield  {author} {\bibinfo {author} {\bibfnamefont {P.~G.}\ \bibnamefont
  {Ortega}}, \bibinfo {author} {\bibfnamefont {J.}~\bibnamefont {Segovia}},
  \bibinfo {author} {\bibfnamefont {D.~R.}\ \bibnamefont {Entem}}, \ and\
  \bibinfo {author} {\bibfnamefont {F.}~\bibnamefont {Fern\'andez}},\ }\href
  {\doibase 10.1103/PhysRevD.94.114018} {\bibfield  {journal} {\bibinfo
  {journal} {Phys. Rev. D}\ }\textbf {\bibinfo {volume} {94}},\ \bibinfo
  {pages} {114018} (\bibinfo {year} {2016})},\ \Eprint
  {http://arxiv.org/abs/1608.01325} {arXiv:1608.01325 [hep-ph]} \BibitemShut
  {NoStop}%
\bibitem [{\citenamefont {Oncala}\ and\ \citenamefont
  {Soto}(2017)}]{Oncala:2017hop}%
  \BibitemOpen
  \bibfield  {author} {\bibinfo {author} {\bibfnamefont {R.}~\bibnamefont
  {Oncala}}\ and\ \bibinfo {author} {\bibfnamefont {J.}~\bibnamefont {Soto}},\
  }\href {\doibase 10.1103/PhysRevD.96.014004} {\bibfield  {journal} {\bibinfo
  {journal} {Phys. Rev. D}\ }\textbf {\bibinfo {volume} {96}},\ \bibinfo
  {pages} {014004} (\bibinfo {year} {2017})},\ \Eprint
  {http://arxiv.org/abs/1702.03900} {arXiv:1702.03900 [hep-ph]} \BibitemShut
  {NoStop}%
\bibitem [{\citenamefont {Ozaki}\ and\ \citenamefont
  {Sasaki}(2013)}]{Ozaki:2012ce}%
  \BibitemOpen
  \bibfield  {author} {\bibinfo {author} {\bibfnamefont {S.}~\bibnamefont
  {Ozaki}}\ and\ \bibinfo {author} {\bibfnamefont {S.}~\bibnamefont {Sasaki}},\
  }\href {\doibase 10.1103/PhysRevD.87.014506} {\bibfield  {journal} {\bibinfo
  {journal} {Phys. Rev. D}\ }\textbf {\bibinfo {volume} {87}},\ \bibinfo
  {pages} {014506} (\bibinfo {year} {2013})},\ \Eprint
  {http://arxiv.org/abs/1211.5512} {arXiv:1211.5512 [hep-lat]} \BibitemShut
  {NoStop}%
\bibitem [{\citenamefont {Panteleeva}\ \emph {et~al.}(2019)\citenamefont
  {Panteleeva}, \citenamefont {Perevalova}, \citenamefont {Polyakov},\ and\
  \citenamefont {Schweitzer}}]{Panteleeva:2018ijz}%
  \BibitemOpen
  \bibfield  {author} {\bibinfo {author} {\bibfnamefont {J.~Y.}\ \bibnamefont
  {Panteleeva}}, \bibinfo {author} {\bibfnamefont {I.~A.}\ \bibnamefont
  {Perevalova}}, \bibinfo {author} {\bibfnamefont {M.~V.}\ \bibnamefont
  {Polyakov}}, \ and\ \bibinfo {author} {\bibfnamefont {P.}~\bibnamefont
  {Schweitzer}},\ }\href {\doibase 10.1103/PhysRevC.99.045206} {\bibfield
  {journal} {\bibinfo  {journal} {Phys. Rev. C}\ }\textbf {\bibinfo {volume}
  {99}},\ \bibinfo {pages} {045206} (\bibinfo {year} {2019})},\ \Eprint
  {http://arxiv.org/abs/1802.09029} {arXiv:1802.09029 [hep-ph]} \BibitemShut
  {NoStop}%
\bibitem [{\citenamefont {Maiani}\ \emph {et~al.}(2016)\citenamefont {Maiani},
  \citenamefont {Polosa},\ and\ \citenamefont {Riquer}}]{Maiani:2016wlq}%
  \BibitemOpen
  \bibfield  {author} {\bibinfo {author} {\bibfnamefont {L.}~\bibnamefont
  {Maiani}}, \bibinfo {author} {\bibfnamefont {A.~D.}\ \bibnamefont {Polosa}},
  \ and\ \bibinfo {author} {\bibfnamefont {V.}~\bibnamefont {Riquer}},\ }\href
  {\doibase 10.1103/PhysRevD.94.054026} {\bibfield  {journal} {\bibinfo
  {journal} {Phys. Rev. D}\ }\textbf {\bibinfo {volume} {94}},\ \bibinfo
  {pages} {054026} (\bibinfo {year} {2016})},\ \Eprint
  {http://arxiv.org/abs/1607.02405} {arXiv:1607.02405 [hep-ph]} \BibitemShut
  {NoStop}%
\bibitem [{\citenamefont {L\"u}\ and\ \citenamefont {Dong}(2016)}]{Lu:2016cwr}%
  \BibitemOpen
  \bibfield  {author} {\bibinfo {author} {\bibfnamefont {Q.-F.}\ \bibnamefont
  {L\"u}}\ and\ \bibinfo {author} {\bibfnamefont {Y.-B.}\ \bibnamefont
  {Dong}},\ }\href {\doibase 10.1103/PhysRevD.94.074007} {\bibfield  {journal}
  {\bibinfo  {journal} {Phys. Rev. D}\ }\textbf {\bibinfo {volume} {94}},\
  \bibinfo {pages} {074007} (\bibinfo {year} {2016})},\ \Eprint
  {http://arxiv.org/abs/1607.05570} {arXiv:1607.05570 [hep-ph]} \BibitemShut
  {NoStop}%
\bibitem [{\citenamefont {Zhu}(2016)}]{Zhu:2016arf}%
  \BibitemOpen
  \bibfield  {author} {\bibinfo {author} {\bibfnamefont {R.}~\bibnamefont
  {Zhu}},\ }\href {\doibase 10.1103/PhysRevD.94.054009} {\bibfield  {journal}
  {\bibinfo  {journal} {Phys. Rev. D}\ }\textbf {\bibinfo {volume} {94}},\
  \bibinfo {pages} {054009} (\bibinfo {year} {2016})},\ \Eprint
  {http://arxiv.org/abs/1607.02799} {arXiv:1607.02799 [hep-ph]} \BibitemShut
  {NoStop}%
\bibitem [{\citenamefont {Deng}\ \emph {et~al.}(2018)\citenamefont {Deng},
  \citenamefont {Ping}, \citenamefont {Huang},\ and\ \citenamefont
  {Wang}}]{Deng:2017xlb}%
  \BibitemOpen
  \bibfield  {author} {\bibinfo {author} {\bibfnamefont {C.}~\bibnamefont
  {Deng}}, \bibinfo {author} {\bibfnamefont {J.}~\bibnamefont {Ping}}, \bibinfo
  {author} {\bibfnamefont {H.}~\bibnamefont {Huang}}, \ and\ \bibinfo {author}
  {\bibfnamefont {F.}~\bibnamefont {Wang}},\ }\href {\doibase
  10.1103/PhysRevD.98.014026} {\bibfield  {journal} {\bibinfo  {journal} {Phys.
  Rev. D}\ }\textbf {\bibinfo {volume} {98}},\ \bibinfo {pages} {014026}
  (\bibinfo {year} {2018})},\ \Eprint {http://arxiv.org/abs/1801.00164}
  {arXiv:1801.00164 [hep-ph]} \BibitemShut {NoStop}%
\bibitem [{\citenamefont {Wu}\ \emph {et~al.}(2016)\citenamefont {Wu},
  \citenamefont {Liu}, \citenamefont {Chen}, \citenamefont {Liu},\ and\
  \citenamefont {Zhu}}]{Wu:2016gas}%
  \BibitemOpen
  \bibfield  {author} {\bibinfo {author} {\bibfnamefont {J.}~\bibnamefont
  {Wu}}, \bibinfo {author} {\bibfnamefont {Y.-R.}\ \bibnamefont {Liu}},
  \bibinfo {author} {\bibfnamefont {K.}~\bibnamefont {Chen}}, \bibinfo {author}
  {\bibfnamefont {X.}~\bibnamefont {Liu}}, \ and\ \bibinfo {author}
  {\bibfnamefont {S.-L.}\ \bibnamefont {Zhu}},\ }\href {\doibase
  10.1103/PhysRevD.94.094031} {\bibfield  {journal} {\bibinfo  {journal} {Phys.
  Rev. D}\ }\textbf {\bibinfo {volume} {94}},\ \bibinfo {pages} {094031}
  (\bibinfo {year} {2016})},\ \Eprint {http://arxiv.org/abs/1608.07900}
  {arXiv:1608.07900 [hep-ph]} \BibitemShut {NoStop}%
\bibitem [{\citenamefont {Yang}\ and\ \citenamefont
  {Ping}(2019)}]{Yang:2019dxd}%
  \BibitemOpen
  \bibfield  {author} {\bibinfo {author} {\bibfnamefont {Y.}~\bibnamefont
  {Yang}}\ and\ \bibinfo {author} {\bibfnamefont {J.}~\bibnamefont {Ping}},\
  }\href {\doibase 10.1103/PhysRevD.99.094032} {\bibfield  {journal} {\bibinfo
  {journal} {Phys. Rev. D}\ }\textbf {\bibinfo {volume} {99}},\ \bibinfo
  {pages} {094032} (\bibinfo {year} {2019})},\ \Eprint
  {http://arxiv.org/abs/1903.08505} {arXiv:1903.08505 [hep-ph]} \BibitemShut
  {NoStop}%
\bibitem [{\citenamefont {Anwar}\ \emph {et~al.}(2018)\citenamefont {Anwar},
  \citenamefont {Ferretti},\ and\ \citenamefont {Santopinto}}]{Anwar:2018sol}%
  \BibitemOpen
  \bibfield  {author} {\bibinfo {author} {\bibfnamefont {M.~N.}\ \bibnamefont
  {Anwar}}, \bibinfo {author} {\bibfnamefont {J.}~\bibnamefont {Ferretti}}, \
  and\ \bibinfo {author} {\bibfnamefont {E.}~\bibnamefont {Santopinto}},\
  }\href {\doibase 10.1103/PhysRevD.98.094015} {\bibfield  {journal} {\bibinfo
  {journal} {Phys. Rev. D}\ }\textbf {\bibinfo {volume} {98}},\ \bibinfo
  {pages} {094015} (\bibinfo {year} {2018})},\ \Eprint
  {http://arxiv.org/abs/1805.06276} {arXiv:1805.06276 [hep-ph]} \BibitemShut
  {NoStop}%
\bibitem [{\citenamefont {Liu}\ \emph {et~al.}(2021)\citenamefont {Liu},
  \citenamefont {Huang}, \citenamefont {Ping}, \citenamefont {Chen},\ and\
  \citenamefont {Zhu}}]{Liu:2021xje}%
  \BibitemOpen
  \bibfield  {author} {\bibinfo {author} {\bibfnamefont {X.}~\bibnamefont
  {Liu}}, \bibinfo {author} {\bibfnamefont {H.}~\bibnamefont {Huang}}, \bibinfo
  {author} {\bibfnamefont {J.}~\bibnamefont {Ping}}, \bibinfo {author}
  {\bibfnamefont {D.}~\bibnamefont {Chen}}, \ and\ \bibinfo {author}
  {\bibfnamefont {X.}~\bibnamefont {Zhu}},\ }\href {\doibase
  10.1140/epjc/s10052-021-09752-y} {\bibfield  {journal} {\bibinfo  {journal}
  {Eur. Phys. J. C}\ }\textbf {\bibinfo {volume} {81}},\ \bibinfo {pages} {950}
  (\bibinfo {year} {2021})},\ \Eprint {http://arxiv.org/abs/2103.12425}
  {arXiv:2103.12425 [hep-ph]} \BibitemShut {NoStop}%
\bibitem [{\citenamefont {Chen}\ \emph {et~al.}(2017)\citenamefont {Chen},
  \citenamefont {Cui}, \citenamefont {Chen}, \citenamefont {Liu},\ and\
  \citenamefont {Zhu}}]{Chen:2016oma}%
  \BibitemOpen
  \bibfield  {author} {\bibinfo {author} {\bibfnamefont {H.-X.}\ \bibnamefont
  {Chen}}, \bibinfo {author} {\bibfnamefont {E.-L.}\ \bibnamefont {Cui}},
  \bibinfo {author} {\bibfnamefont {W.}~\bibnamefont {Chen}}, \bibinfo {author}
  {\bibfnamefont {X.}~\bibnamefont {Liu}}, \ and\ \bibinfo {author}
  {\bibfnamefont {S.-L.}\ \bibnamefont {Zhu}},\ }\href {\doibase
  10.1140/epjc/s10052-017-4737-5} {\bibfield  {journal} {\bibinfo  {journal}
  {Eur. Phys. J. C}\ }\textbf {\bibinfo {volume} {77}},\ \bibinfo {pages} {160}
  (\bibinfo {year} {2017})},\ \Eprint {http://arxiv.org/abs/1606.03179}
  {arXiv:1606.03179 [hep-ph]} \BibitemShut {NoStop}%
\bibitem [{\citenamefont {Wang}(2017)}]{Wang:2016gxp}%
  \BibitemOpen
  \bibfield  {author} {\bibinfo {author} {\bibfnamefont {Z.-G.}\ \bibnamefont
  {Wang}},\ }\href {\doibase 10.1140/epjc/s10052-017-4640-0} {\bibfield
  {journal} {\bibinfo  {journal} {Eur. Phys. J. C}\ }\textbf {\bibinfo {volume}
  {77}},\ \bibinfo {pages} {78} (\bibinfo {year} {2017})},\ \Eprint
  {http://arxiv.org/abs/1606.05872} {arXiv:1606.05872 [hep-ph]} \BibitemShut
  {NoStop}%
\bibitem [{\citenamefont {Reinders}\ \emph {et~al.}(1985)\citenamefont
  {Reinders}, \citenamefont {Rubinstein},\ and\ \citenamefont
  {Yazaki}}]{Reinders:1984sr}%
  \BibitemOpen
  \bibfield  {author} {\bibinfo {author} {\bibfnamefont {L.~J.}\ \bibnamefont
  {Reinders}}, \bibinfo {author} {\bibfnamefont {H.}~\bibnamefont
  {Rubinstein}}, \ and\ \bibinfo {author} {\bibfnamefont {S.}~\bibnamefont
  {Yazaki}},\ }\href {\doibase 10.1016/0370-1573(85)90065-1} {\bibfield
  {journal} {\bibinfo  {journal} {Phys. Rept.}\ }\textbf {\bibinfo {volume}
  {127}},\ \bibinfo {pages} {1} (\bibinfo {year} {1985})}\BibitemShut {NoStop}%
\bibitem [{\citenamefont {Nakamura}(2021)}]{Nakamura:2021bvs}%
  \BibitemOpen
  \bibfield  {author} {\bibinfo {author} {\bibfnamefont {S.~X.}\ \bibnamefont
  {Nakamura}},\ }\href@noop {} {\  (\bibinfo {year} {2021})},\ \Eprint
  {http://arxiv.org/abs/2111.05115} {arXiv:2111.05115 [hep-ph]} \BibitemShut
  {NoStop}%
\bibitem [{\citenamefont {Liu}(2017)}]{Liu:2016onn}%
  \BibitemOpen
  \bibfield  {author} {\bibinfo {author} {\bibfnamefont {X.-H.}\ \bibnamefont
  {Liu}},\ }\href {\doibase 10.1016/j.physletb.2017.01.008} {\bibfield
  {journal} {\bibinfo  {journal} {Phys. Lett. B}\ }\textbf {\bibinfo {volume}
  {766}},\ \bibinfo {pages} {117} (\bibinfo {year} {2017})},\ \Eprint
  {http://arxiv.org/abs/1607.01385} {arXiv:1607.01385 [hep-ph]} \BibitemShut
  {NoStop}%
\bibitem [{\citenamefont {Ge}\ \emph {et~al.}(2021)\citenamefont {Ge},
  \citenamefont {Liu},\ and\ \citenamefont {Ke}}]{Ge:2021sdq}%
  \BibitemOpen
  \bibfield  {author} {\bibinfo {author} {\bibfnamefont {Y.-H.}\ \bibnamefont
  {Ge}}, \bibinfo {author} {\bibfnamefont {X.-H.}\ \bibnamefont {Liu}}, \ and\
  \bibinfo {author} {\bibfnamefont {H.-W.}\ \bibnamefont {Ke}},\ }\href
  {\doibase 10.1140/epjc/s10052-021-09590-y} {\  (\bibinfo {year} {2021}),\
  10.1140/epjc/s10052-021-09590-y},\ \Eprint {http://arxiv.org/abs/2103.05282}
  {arXiv:2103.05282 [hep-ph]} \BibitemShut {NoStop}%
\bibitem [{\citenamefont {Zyla}\ \emph {et~al.}(2020)\citenamefont {Zyla} \emph
  {et~al.}}]{ParticleDataGroup:2020ssz}%
  \BibitemOpen
  \bibfield  {author} {\bibinfo {author} {\bibfnamefont {P.~A.}\ \bibnamefont
  {Zyla}} \emph {et~al.} (\bibinfo {collaboration} {Particle Data Group}),\
  }\href {\doibase 10.1093/ptep/ptaa104} {\bibfield  {journal} {\bibinfo
  {journal} {PTEP}\ }\textbf {\bibinfo {volume} {2020}},\ \bibinfo {pages}
  {083C01} (\bibinfo {year} {2020})}\BibitemShut {NoStop}%
\bibitem [{\citenamefont {Albuquerque}\ and\ \citenamefont
  {Nielsen}(2009)}]{Albuquerque:2008up}%
  \BibitemOpen
  \bibfield  {author} {\bibinfo {author} {\bibfnamefont {R.~M.}\ \bibnamefont
  {Albuquerque}}\ and\ \bibinfo {author} {\bibfnamefont {M.}~\bibnamefont
  {Nielsen}},\ }\href {\doibase 10.1016/j.nuclphysa.2011.04.001} {\bibfield
  {journal} {\bibinfo  {journal} {Nucl. Phys. A}\ }\textbf {\bibinfo {volume}
  {815}},\ \bibinfo {pages} {53} (\bibinfo {year} {2009})},\ \bibinfo {note}
  {[Erratum: Nucl.Phys.A 857, 48--49 (2011)]},\ \Eprint
  {http://arxiv.org/abs/0804.4817} {arXiv:0804.4817 [hep-ph]} \BibitemShut
  {NoStop}%
\bibitem [{\citenamefont {Be\v{c}irevi\'c}\ \emph {et~al.}(2014)\citenamefont
  {Be\v{c}irevi\'c}, \citenamefont {Duplan\v{c}i\'c}, \citenamefont {Klajn},
  \citenamefont {Meli\'c},\ and\ \citenamefont
  {Sanfilippo}}]{Becirevic:2013bsa}%
  \BibitemOpen
  \bibfield  {author} {\bibinfo {author} {\bibfnamefont {D.}~\bibnamefont
  {Be\v{c}irevi\'c}}, \bibinfo {author} {\bibfnamefont {G.}~\bibnamefont
  {Duplan\v{c}i\'c}}, \bibinfo {author} {\bibfnamefont {B.}~\bibnamefont
  {Klajn}}, \bibinfo {author} {\bibfnamefont {B.}~\bibnamefont {Meli\'c}}, \
  and\ \bibinfo {author} {\bibfnamefont {F.}~\bibnamefont {Sanfilippo}},\
  }\href {\doibase 10.1016/j.nuclphysb.2014.03.024} {\bibfield  {journal}
  {\bibinfo  {journal} {Nucl. Phys. B}\ }\textbf {\bibinfo {volume} {883}},\
  \bibinfo {pages} {306} (\bibinfo {year} {2014})},\ \Eprint
  {http://arxiv.org/abs/1312.2858} {arXiv:1312.2858 [hep-ph]} \BibitemShut
  {NoStop}%
\bibitem [{\citenamefont {Belyaev}\ \emph {et~al.}(1995)\citenamefont
  {Belyaev}, \citenamefont {Braun}, \citenamefont {Khodjamirian},\ and\
  \citenamefont {Ruckl}}]{Belyaev:1994zk}%
  \BibitemOpen
  \bibfield  {author} {\bibinfo {author} {\bibfnamefont {V.~M.}\ \bibnamefont
  {Belyaev}}, \bibinfo {author} {\bibfnamefont {V.~M.}\ \bibnamefont {Braun}},
  \bibinfo {author} {\bibfnamefont {A.}~\bibnamefont {Khodjamirian}}, \ and\
  \bibinfo {author} {\bibfnamefont {R.}~\bibnamefont {Ruckl}},\ }\href
  {\doibase 10.1103/PhysRevD.51.6177} {\bibfield  {journal} {\bibinfo
  {journal} {Phys. Rev. D}\ }\textbf {\bibinfo {volume} {51}},\ \bibinfo
  {pages} {6177} (\bibinfo {year} {1995})},\ \Eprint
  {http://arxiv.org/abs/hep-ph/9410280} {arXiv:hep-ph/9410280} \BibitemShut
  {NoStop}%
\bibitem [{\citenamefont {Ioffe}\ and\ \citenamefont
  {Smilga}(1984)}]{Ioffe:1983ju}%
  \BibitemOpen
  \bibfield  {author} {\bibinfo {author} {\bibfnamefont {B.~L.}\ \bibnamefont
  {Ioffe}}\ and\ \bibinfo {author} {\bibfnamefont {A.~V.}\ \bibnamefont
  {Smilga}},\ }\href {\doibase 10.1016/0550-3213(84)90364-X} {\bibfield
  {journal} {\bibinfo  {journal} {Nucl. Phys. B}\ }\textbf {\bibinfo {volume}
  {232}},\ \bibinfo {pages} {109} (\bibinfo {year} {1984})}\BibitemShut
  {NoStop}%
\bibitem [{\citenamefont {Dias}\ \emph {et~al.}(2013)\citenamefont {Dias},
  \citenamefont {Navarra}, \citenamefont {Nielsen},\ and\ \citenamefont
  {Zanetti}}]{Dias:2013xfa}%
  \BibitemOpen
  \bibfield  {author} {\bibinfo {author} {\bibfnamefont {J.~M.}\ \bibnamefont
  {Dias}}, \bibinfo {author} {\bibfnamefont {F.~S.}\ \bibnamefont {Navarra}},
  \bibinfo {author} {\bibfnamefont {M.}~\bibnamefont {Nielsen}}, \ and\
  \bibinfo {author} {\bibfnamefont {C.~M.}\ \bibnamefont {Zanetti}},\ }\href
  {\doibase 10.1103/PhysRevD.88.016004} {\bibfield  {journal} {\bibinfo
  {journal} {Phys. Rev. D}\ }\textbf {\bibinfo {volume} {88}},\ \bibinfo
  {pages} {016004} (\bibinfo {year} {2013})},\ \Eprint
  {http://arxiv.org/abs/1304.6433} {arXiv:1304.6433 [hep-ph]} \BibitemShut
  {NoStop}%
\bibitem [{\citenamefont {Huang}\ \emph {et~al.}(2011)\citenamefont {Huang},
  \citenamefont {Chen},\ and\ \citenamefont {Zhu}}]{Huang:2010dc}%
  \BibitemOpen
  \bibfield  {author} {\bibinfo {author} {\bibfnamefont {P.-Z.}\ \bibnamefont
  {Huang}}, \bibinfo {author} {\bibfnamefont {H.-X.}\ \bibnamefont {Chen}}, \
  and\ \bibinfo {author} {\bibfnamefont {S.-L.}\ \bibnamefont {Zhu}},\ }\href
  {\doibase 10.1103/PhysRevD.83.014021} {\bibfield  {journal} {\bibinfo
  {journal} {Phys. Rev. D}\ }\textbf {\bibinfo {volume} {83}},\ \bibinfo
  {pages} {014021} (\bibinfo {year} {2011})},\ \Eprint
  {http://arxiv.org/abs/1010.2293} {arXiv:1010.2293 [hep-ph]} \BibitemShut
  {NoStop}%
\bibitem [{\citenamefont {Agaev}\ \emph
  {et~al.}(2016{\natexlab{a}})\citenamefont {Agaev}, \citenamefont {Azizi},\
  and\ \citenamefont {Sundu}}]{Agaev:2016mjb}%
  \BibitemOpen
  \bibfield  {author} {\bibinfo {author} {\bibfnamefont {S.~S.}\ \bibnamefont
  {Agaev}}, \bibinfo {author} {\bibfnamefont {K.}~\bibnamefont {Azizi}}, \ and\
  \bibinfo {author} {\bibfnamefont {H.}~\bibnamefont {Sundu}},\ }\href
  {\doibase 10.1103/PhysRevD.93.074024} {\bibfield  {journal} {\bibinfo
  {journal} {Phys. Rev. D}\ }\textbf {\bibinfo {volume} {93}},\ \bibinfo
  {pages} {074024} (\bibinfo {year} {2016}{\natexlab{a}})},\ \Eprint
  {http://arxiv.org/abs/1602.08642} {arXiv:1602.08642 [hep-ph]} \BibitemShut
  {NoStop}%
\bibitem [{\citenamefont {Agaev}\ \emph
  {et~al.}(2016{\natexlab{b}})\citenamefont {Agaev}, \citenamefont {Azizi},\
  and\ \citenamefont {Sundu}}]{Agaev:2016dev}%
  \BibitemOpen
  \bibfield  {author} {\bibinfo {author} {\bibfnamefont {S.~S.}\ \bibnamefont
  {Agaev}}, \bibinfo {author} {\bibfnamefont {K.}~\bibnamefont {Azizi}}, \ and\
  \bibinfo {author} {\bibfnamefont {H.}~\bibnamefont {Sundu}},\ }\href
  {\doibase 10.1103/PhysRevD.93.074002} {\bibfield  {journal} {\bibinfo
  {journal} {Phys. Rev. D}\ }\textbf {\bibinfo {volume} {93}},\ \bibinfo
  {pages} {074002} (\bibinfo {year} {2016}{\natexlab{b}})},\ \Eprint
  {http://arxiv.org/abs/1601.03847} {arXiv:1601.03847 [hep-ph]} \BibitemShut
  {NoStop}%
\bibitem [{\citenamefont {Matheus}\ \emph {et~al.}(2007)\citenamefont
  {Matheus}, \citenamefont {Narison}, \citenamefont {Nielsen},\ and\
  \citenamefont {Richard}}]{Matheus:2006xi}%
  \BibitemOpen
  \bibfield  {author} {\bibinfo {author} {\bibfnamefont {R.~D.}\ \bibnamefont
  {Matheus}}, \bibinfo {author} {\bibfnamefont {S.}~\bibnamefont {Narison}},
  \bibinfo {author} {\bibfnamefont {M.}~\bibnamefont {Nielsen}}, \ and\
  \bibinfo {author} {\bibfnamefont {J.~M.}\ \bibnamefont {Richard}},\ }\href
  {\doibase 10.1103/PhysRevD.75.014005} {\bibfield  {journal} {\bibinfo
  {journal} {Phys. Rev. D}\ }\textbf {\bibinfo {volume} {75}},\ \bibinfo
  {pages} {014005} (\bibinfo {year} {2007})},\ \Eprint
  {http://arxiv.org/abs/hep-ph/0608297} {arXiv:hep-ph/0608297} \BibitemShut
  {NoStop}%
\bibitem [{\citenamefont {Colangelo}\ \emph {et~al.}(2010)\citenamefont
  {Colangelo}, \citenamefont {De~Fazio},\ and\ \citenamefont
  {Wang}}]{Colangelo:2010bg}%
  \BibitemOpen
  \bibfield  {author} {\bibinfo {author} {\bibfnamefont {P.}~\bibnamefont
  {Colangelo}}, \bibinfo {author} {\bibfnamefont {F.}~\bibnamefont {De~Fazio}},
  \ and\ \bibinfo {author} {\bibfnamefont {W.}~\bibnamefont {Wang}},\ }\href
  {\doibase 10.1103/PhysRevD.81.074001} {\bibfield  {journal} {\bibinfo
  {journal} {Phys. Rev. D}\ }\textbf {\bibinfo {volume} {81}},\ \bibinfo
  {pages} {074001} (\bibinfo {year} {2010})},\ \Eprint
  {http://arxiv.org/abs/1002.2880} {arXiv:1002.2880 [hep-ph]} \BibitemShut
  {NoStop}%
\bibitem [{\citenamefont {Ball}\ \emph {et~al.}(2007)\citenamefont {Ball},
  \citenamefont {Braun},\ and\ \citenamefont {Lenz}}]{Ball:2007zt}%
  \BibitemOpen
  \bibfield  {author} {\bibinfo {author} {\bibfnamefont {P.}~\bibnamefont
  {Ball}}, \bibinfo {author} {\bibfnamefont {V.~M.}\ \bibnamefont {Braun}}, \
  and\ \bibinfo {author} {\bibfnamefont {A.}~\bibnamefont {Lenz}},\ }\href
  {\doibase 10.1088/1126-6708/2007/08/090} {\bibfield  {journal} {\bibinfo
  {journal} {JHEP}\ }\textbf {\bibinfo {volume} {08}},\ \bibinfo {pages} {090}
  (\bibinfo {year} {2007})},\ \Eprint {http://arxiv.org/abs/0707.1201}
  {arXiv:0707.1201 [hep-ph]} \BibitemShut {NoStop}%
\bibitem [{\citenamefont {Hu}\ \emph {et~al.}(2021)\citenamefont {Hu},
  \citenamefont {Fu}, \citenamefont {Zhong}, \citenamefont {Wu},\ and\
  \citenamefont {Wu}}]{Hu:2021lkl}%
  \BibitemOpen
  \bibfield  {author} {\bibinfo {author} {\bibfnamefont {D.-D.}\ \bibnamefont
  {Hu}}, \bibinfo {author} {\bibfnamefont {H.-B.}\ \bibnamefont {Fu}}, \bibinfo
  {author} {\bibfnamefont {T.}~\bibnamefont {Zhong}}, \bibinfo {author}
  {\bibfnamefont {Z.-H.}\ \bibnamefont {Wu}}, \ and\ \bibinfo {author}
  {\bibfnamefont {X.-G.}\ \bibnamefont {Wu}},\ }\href@noop {} {\  (\bibinfo
  {year} {2021})},\ \Eprint {http://arxiv.org/abs/2107.02758} {arXiv:2107.02758
  [hep-ph]} \BibitemShut {NoStop}%
\bibitem [{\citenamefont {Olpak}\ \emph {et~al.}(2017)\citenamefont {Olpak},
  \citenamefont {Ozpineci},\ and\ \citenamefont {Tanriverdi}}]{Olpak:2016wkf}%
  \BibitemOpen
  \bibfield  {author} {\bibinfo {author} {\bibfnamefont {M.~A.}\ \bibnamefont
  {Olpak}}, \bibinfo {author} {\bibfnamefont {A.}~\bibnamefont {Ozpineci}}, \
  and\ \bibinfo {author} {\bibfnamefont {V.}~\bibnamefont {Tanriverdi}},\
  }\href {\doibase 10.1103/PhysRevD.96.014026} {\bibfield  {journal} {\bibinfo
  {journal} {Phys. Rev. D}\ }\textbf {\bibinfo {volume} {96}},\ \bibinfo
  {pages} {014026} (\bibinfo {year} {2017})},\ \Eprint
  {http://arxiv.org/abs/1608.04539} {arXiv:1608.04539 [hep-ph]} \BibitemShut
  {NoStop}%
\bibitem [{\citenamefont {Beringer}\ \emph {et~al.}(2012)\citenamefont
  {Beringer} \emph {et~al.}}]{ParticleDataGroup:2012pjm}%
  \BibitemOpen
  \bibfield  {author} {\bibinfo {author} {\bibfnamefont {J.}~\bibnamefont
  {Beringer}} \emph {et~al.} (\bibinfo {collaboration} {Particle Data Group}),\
  }\href {\doibase 10.1103/PhysRevD.86.010001} {\bibfield  {journal} {\bibinfo
  {journal} {Phys. Rev. D}\ }\textbf {\bibinfo {volume} {86}},\ \bibinfo
  {pages} {010001} (\bibinfo {year} {2012})}\BibitemShut {NoStop}%
\bibitem [{\citenamefont {Wang}\ and\ \citenamefont
  {Xin}(2021)}]{Wang:2021lkg}%
  \BibitemOpen
  \bibfield  {author} {\bibinfo {author} {\bibfnamefont {Z.-G.}\ \bibnamefont
  {Wang}}\ and\ \bibinfo {author} {\bibfnamefont {Q.}~\bibnamefont {Xin}},\
  }\href@noop {} {\  (\bibinfo {year} {2021})},\ \Eprint
  {http://arxiv.org/abs/2112.04776} {arXiv:2112.04776 [hep-ph]} \BibitemShut
  {NoStop}%
\bibitem [{\citenamefont {Wang}(2021)}]{Wang:2021ghk}%
  \BibitemOpen
  \bibfield  {author} {\bibinfo {author} {\bibfnamefont {Z.-G.}\ \bibnamefont
  {Wang}},\ }\href {\doibase 10.1155/2021/4426163} {\bibfield  {journal}
  {\bibinfo  {journal} {Adv. High Energy Phys.}\ }\textbf {\bibinfo {volume}
  {2021}},\ \bibinfo {pages} {4426163} (\bibinfo {year} {2021})},\ \Eprint
  {http://arxiv.org/abs/2103.04236} {arXiv:2103.04236 [hep-ph]} \BibitemShut
  {NoStop}%
\bibitem [{\citenamefont {Wang}\ and\ \citenamefont {Di}(2019)}]{Wang:2018qpe}%
  \BibitemOpen
  \bibfield  {author} {\bibinfo {author} {\bibfnamefont {Z.-G.}\ \bibnamefont
  {Wang}}\ and\ \bibinfo {author} {\bibfnamefont {Z.-Y.}\ \bibnamefont {Di}},\
  }\href {\doibase 10.1140/epjc/s10052-019-6596-8} {\bibfield  {journal}
  {\bibinfo  {journal} {Eur. Phys. J. C}\ }\textbf {\bibinfo {volume} {79}},\
  \bibinfo {pages} {72} (\bibinfo {year} {2019})},\ \Eprint
  {http://arxiv.org/abs/1811.12821} {arXiv:1811.12821 [hep-ph]} \BibitemShut
  {NoStop}%
\bibitem [{\citenamefont {Maiani}\ \emph {et~al.}(2014)\citenamefont {Maiani},
  \citenamefont {Piccinini}, \citenamefont {Polosa},\ and\ \citenamefont
  {Riquer}}]{Maiani:2014aja}%
  \BibitemOpen
  \bibfield  {author} {\bibinfo {author} {\bibfnamefont {L.}~\bibnamefont
  {Maiani}}, \bibinfo {author} {\bibfnamefont {F.}~\bibnamefont {Piccinini}},
  \bibinfo {author} {\bibfnamefont {A.~D.}\ \bibnamefont {Polosa}}, \ and\
  \bibinfo {author} {\bibfnamefont {V.}~\bibnamefont {Riquer}},\ }\href
  {\doibase 10.1103/PhysRevD.89.114010} {\bibfield  {journal} {\bibinfo
  {journal} {Phys. Rev. D}\ }\textbf {\bibinfo {volume} {89}},\ \bibinfo
  {pages} {114010} (\bibinfo {year} {2014})},\ \Eprint
  {http://arxiv.org/abs/1405.1551} {arXiv:1405.1551 [hep-ph]} \BibitemShut
  {NoStop}%
\bibitem [{\citenamefont {Nielsen}\ and\ \citenamefont
  {Navarra}(2014)}]{Nielsen:2014mva}%
  \BibitemOpen
  \bibfield  {author} {\bibinfo {author} {\bibfnamefont {M.}~\bibnamefont
  {Nielsen}}\ and\ \bibinfo {author} {\bibfnamefont {F.~S.}\ \bibnamefont
  {Navarra}},\ }\href {\doibase 10.1142/S0217732314300055} {\bibfield
  {journal} {\bibinfo  {journal} {Mod. Phys. Lett. A}\ }\textbf {\bibinfo
  {volume} {29}},\ \bibinfo {pages} {1430005} (\bibinfo {year} {2014})},\
  \Eprint {http://arxiv.org/abs/1401.2913} {arXiv:1401.2913 [hep-ph]}
  \BibitemShut {NoStop}%
\bibitem [{\citenamefont {Wang}(2015)}]{Wang:2014vha}%
  \BibitemOpen
  \bibfield  {author} {\bibinfo {author} {\bibfnamefont {Z.-G.}\ \bibnamefont
  {Wang}},\ }\href {\doibase 10.1088/0253-6102/63/3/325} {\bibfield  {journal}
  {\bibinfo  {journal} {Commun. Theor. Phys.}\ }\textbf {\bibinfo {volume}
  {63}},\ \bibinfo {pages} {325} (\bibinfo {year} {2015})},\ \Eprint
  {http://arxiv.org/abs/1405.3581} {arXiv:1405.3581 [hep-ph]} \BibitemShut
  {NoStop}%
\bibitem [{\citenamefont {Lebed}\ and\ \citenamefont
  {Polosa}(2016)}]{Lebed:2016yvr}%
  \BibitemOpen
  \bibfield  {author} {\bibinfo {author} {\bibfnamefont {R.~F.}\ \bibnamefont
  {Lebed}}\ and\ \bibinfo {author} {\bibfnamefont {A.~D.}\ \bibnamefont
  {Polosa}},\ }\href {\doibase 10.1103/PhysRevD.93.094024} {\bibfield
  {journal} {\bibinfo  {journal} {Phys. Rev. D}\ }\textbf {\bibinfo {volume}
  {93}},\ \bibinfo {pages} {094024} (\bibinfo {year} {2016})},\ \Eprint
  {http://arxiv.org/abs/1602.08421} {arXiv:1602.08421 [hep-ph]} \BibitemShut
  {NoStop}%
\bibitem [{\citenamefont {Chen}\ and\ \citenamefont
  {Chen}(2019)}]{Chen:2019osl}%
  \BibitemOpen
  \bibfield  {author} {\bibinfo {author} {\bibfnamefont {H.-X.}\ \bibnamefont
  {Chen}}\ and\ \bibinfo {author} {\bibfnamefont {W.}~\bibnamefont {Chen}},\
  }\href {\doibase 10.1103/PhysRevD.99.074022} {\bibfield  {journal} {\bibinfo
  {journal} {Phys. Rev. D}\ }\textbf {\bibinfo {volume} {99}},\ \bibinfo
  {pages} {074022} (\bibinfo {year} {2019})},\ \Eprint
  {http://arxiv.org/abs/1901.06946} {arXiv:1901.06946 [hep-ph]} \BibitemShut
  {NoStop}%
\bibitem [{\citenamefont {Wang}(2020)}]{Wang:2019hnw}%
  \BibitemOpen
  \bibfield  {author} {\bibinfo {author} {\bibfnamefont {Z.-G.}\ \bibnamefont
  {Wang}},\ }\href {\doibase 10.1088/1674-1137/44/6/063105} {\bibfield
  {journal} {\bibinfo  {journal} {Chin. Phys. C}\ }\textbf {\bibinfo {volume}
  {44}},\ \bibinfo {pages} {063105} (\bibinfo {year} {2020})},\ \Eprint
  {http://arxiv.org/abs/1901.10741} {arXiv:1901.10741 [hep-ph]} \BibitemShut
  {NoStop}%
\bibitem [{\citenamefont {Ablikim}\ \emph {et~al.}(2017)\citenamefont {Ablikim}
  \emph {et~al.}}]{BESIII:2016bnd}%
  \BibitemOpen
  \bibfield  {author} {\bibinfo {author} {\bibfnamefont {M.}~\bibnamefont
  {Ablikim}} \emph {et~al.} (\bibinfo {collaboration} {BESIII}),\ }\href
  {\doibase 10.1103/PhysRevLett.118.092001} {\bibfield  {journal} {\bibinfo
  {journal} {Phys. Rev. Lett}\ }\textbf {\bibinfo {volume} {118}},\ \bibinfo
  {pages} {092001} (\bibinfo {year} {2017})},\ \Eprint
  {http://arxiv.org/abs/1611.01317} {arXiv:1611.01317 [hep-ex]} \BibitemShut
  {NoStop}%
\bibitem [{\citenamefont {Ball}\ and\ \citenamefont
  {Braun}(1996)}]{Ball:1996tb}%
  \BibitemOpen
  \bibfield  {author} {\bibinfo {author} {\bibfnamefont {P.}~\bibnamefont
  {Ball}}\ and\ \bibinfo {author} {\bibfnamefont {V.~M.}\ \bibnamefont
  {Braun}},\ }\href {\doibase 10.1103/PhysRevD.54.2182} {\bibfield  {journal}
  {\bibinfo  {journal} {Phys. Rev. D}\ }\textbf {\bibinfo {volume} {54}},\
  \bibinfo {pages} {2182} (\bibinfo {year} {1996})},\ \Eprint
  {http://arxiv.org/abs/hep-ph/9602323} {arXiv:hep-ph/9602323} \BibitemShut
  {NoStop}%
\bibitem [{\citenamefont {Ball}\ and\ \citenamefont
  {Jones}(2007)}]{Ball:2007rt}%
  \BibitemOpen
  \bibfield  {author} {\bibinfo {author} {\bibfnamefont {P.}~\bibnamefont
  {Ball}}\ and\ \bibinfo {author} {\bibfnamefont {G.~W.}\ \bibnamefont
  {Jones}},\ }\href {\doibase 10.1088/1126-6708/2007/03/069} {\bibfield
  {journal} {\bibinfo  {journal} {JHEP}\ }\textbf {\bibinfo {volume} {03}},\
  \bibinfo {pages} {069} (\bibinfo {year} {2007})},\ \Eprint
  {http://arxiv.org/abs/hep-ph/0702100} {arXiv:hep-ph/0702100} \BibitemShut
  {NoStop}%
\bibitem [{\citenamefont {Gubler}(2013)}]{Gubler:2013moa}%
  \BibitemOpen
  \bibfield  {author} {\bibinfo {author} {\bibfnamefont {P.}~\bibnamefont
  {Gubler}},\ }\emph {\bibinfo {title} {{A Bayesian Analysis of QCD Sum
  Rules}}},\ \href {\doibase 10.1007/978-4-431-54318-3} {Ph.D. thesis},\
  \bibinfo  {school} {Tokyo Inst. Tech.}, \bibinfo {address} {Tokyo} (\bibinfo
  {year} {2013})\BibitemShut {NoStop}%
\end{thebibliography}%

\end{document}